\documentclass[conference]{IEEEtran}
\IEEEoverridecommandlockouts
\usepackage{cite}
\usepackage{amsmath,amssymb,amsfonts}
\usepackage{algorithmic}
\usepackage{graphicx}
\usepackage{textcomp}
\usepackage{xcolor}
\def\BibTeX{{\rm B\kern-.05em{\sc i\kern-.025em b}\kern-.08em
    T\kern-.1667em\lower.7ex\hbox{E}\kern-.125emX}}
\usepackage{braket}
\usepackage{soul}
\usepackage{subcaption}
\usepackage{float}

\usepackage[most]{tcolorbox}

\newcounter{obscounter}
\renewcommand{\theobscounter}{\Roman{obscounter}}
\definecolor{custompurple}{HTML}{53257F}
\newtcolorbox{ObservationBox}[2][]{text width=0.94\linewidth,
colbacktitle=custompurple,enhanced,
attach boxed title to top left={yshift=-2mm,xshift=3mm},
boxed title style={sharp corners},top=6pt,bottom=2pt,
title=#2,colback=custompurple!10!white, left=4pt, right=2pt}
\newcommand{\obs}[1]{
\refstepcounter{obscounter}
\begin{ObservationBox}{\textbf{Observation \theobscounter}}
\par\noindent
#1
\end{ObservationBox}}

\tcbset{textmarker/.style={%
        enhanced,
        parbox=false,boxrule=0mm,boxsep=-1mm,arc=0mm,
        outer arc=0mm,left=3mm,right=3mm,top=7pt,bottom=7pt,
        toptitle=1mm,bottomtitle=1mm,oversize}}
\newtcolorbox{noteBox}{textmarker,
    borderline west={3pt}{0pt}{red},
    colback=red!10!white}

\begin{document}

\title{On the Efficacy of Surface Codes in Compensating for Radiation Events in Superconducting Devices}

\author{\IEEEauthorblockN{1\textsuperscript{st} Marzio Vallero}
\IEEEauthorblockA{
\textit{University of Trento}\\
Trento, Italy \\
marzio.vallero@unitn.it}
\and
\IEEEauthorblockN{2\textsuperscript{nd} Gioele Casagranda}
\IEEEauthorblockA{
\textit{University of Trento}\\
Trento, Italy \\
gioele.casagranda@unitn.it}
\and
\IEEEauthorblockN{3\textsuperscript{rd} Flavio Vella}
\IEEEauthorblockA{
\textit{University of Trento}\\
Trento, Italy \\
flavio.vella@unitn.it}
\and
\IEEEauthorblockN{4\textsuperscript{th} Paolo Rech}
\IEEEauthorblockA{
\textit{University of Trento}\\
Trento, Italy \\
paolo.rech@unitn.it}
}

\maketitle

\begin{abstract}
    Reliability is fundamental for developing large-scale quantum computers. Since the benefit of technological advancements to the qubit's stability is saturating, algorithmic solutions, such as quantum error correction (QEC) codes, are needed to bridge the gap to reliable computation.
    Unfortunately, the deployment of the first quantum computers has identified faults induced by natural radiation as an additional threat to qubits reliability.
    The high sensitivity of qubits to radiation hinders the large-scale adoption of quantum computers, since the persistence and area-of-effect of the fault can potentially undermine the efficacy of the most advanced QEC.

    In this paper, we investigate the resilience of various implementations of state-of-the-art QEC codes to radiation-induced faults. We report data from over 400 million fault injections and correlate hardware faults with the logical error observed after decoding the code output, extrapolating physical-to-logical error rates.
    We compare the code's radiation-induced logical error rate over the code distance, the number and role in the QEC of physical qubits, the underlying quantum computer topology, and particle energy spread in the chip.
    We show that, by simply selecting and tuning properly the surface code, thus without introducing any overhead, the probability of correcting a radiation-induced fault is increased by up to 10\%.
    Finally, we provide indications and guidelines for the design of future QEC codes to further increase their effectiveness against radiation-induced events.

\end{abstract}


\section{Introduction}
Ever since its theoretical inception, 
quantum computing (QC) has promised unprecedented improvements in computation efficiency and time to solution for both computationally hard and classically intractable problems \cite{shor1994factorisation, 1996Grover}.
Despite the tremendous technological advancements in recent years, quantum hardware designers are still facing issues in merging the reliability and scalability aspects of quantum computers\cite{preskill1998reliability, vanmeter2016, sete2016, copsey2003}.
Significant improvements are still necessary to fill the gap between current Noisy Intermediate Scale Quantum (NISQ) devices and large-scale fault-tolerant quantum machines.

There are various reliability challenges associated with quantum computing, either intrinsic to the implementation of quantum bits (qubits) 
or caused by the inevitable interaction with the environment~\cite{Unruh1995, divincenzo1999coherence}.
Current advanced qubit implementations grant a quantum state stability slowly transitioning towards coherence times greater than $1.4$ \textit{ms} \cite{Stassi2020, Wang2022, Somoroff2023}, and thanks to special gate composition and pulse scheduling techniques, quantum gate fidelity now ranges from $\sim 85\%$ to upwards of $\sim 99\%$ \cite{Ghosh2013, Willsch2017, Rol2019, Kim2022, AbuGhanem2024}. 

Technological improvements, however, are reaching their limits and any further upgrade is bound by design and development costs~\cite{Gambetta2017, Wendin2017, Vischi2022}.
Quantum circuit designers have thus developed quantum error correction (QEC) strategies to mitigate hardware faults in software.
 Various physical qubits are intertwined in a \textit{surface code} and their properties are then combined to encode a single \textit{logical} qubit.
QEC approaches 
have 
lowered 
the error rate of quantum devices~\cite{JavadiAbhari2017, Andersen2020, Krinner2022, Goto2023, Tiurev2023correcting, Katsuda2024}
but require high resource cost: 
multiple physical qubits are needed to encode a single quantum logical qubit, with an overhead going from $7\times$ \cite{Steane1996} to upwards of $49\times$ \cite{Bonilla_Ataides_2021, Tiurev2023correcting, Katsuda2024}.

Unfortunately, the deployment of the first quantum computers has highlighted an incredibly high susceptibility of superconducting qubits to radiation~\cite{Barends2011, radiation2011, LossMechanisms2018, nature_rad, Wilen2021, Cardani2021, Martinis2021, Chen2021, Acharya2023}. 
\textit{Any} particle interaction can alter the qubit(s) state, forcing them into a decoherent state for long periods of time (up to 100s of seconds)~\cite{Oliveira2023neutrons}. 
The rate of occurrence of radiation events in qubits has been measured to be every \textit{tens of seconds}~\cite{Acharya2023}, several orders of magnitude greater than in CMOS transistors~\cite{Oliveira2017}. 
Crucially, while significant effort has been made to reduce the radiation impact in classical computation, the transient error detection/correction in quantum circuits is largely unexplored.

Given the advances in software QEC and the available knowledge about radiation corruption in quantum chips, this work aims to answer the following research questions:

\begin{itemize}
    \item \textbf{RQ1}: Are state-of-the-art error correction codes, designed for intrinsic noise, effective for radiation-induced events?
    \item \textbf{RQ2}: How can we tune and configure the surface code to improve the chance of correcting also radiation strikes?
    \item \textbf{RQ3}: Which are the main insights to design future reliability solutions for radiation-induced corruptions?
\end{itemize}

To provide a realistic evaluation, we modelled the radiation-induced fault following its theoretical definition and experimental observation, considering both the fault's temporal and spatial distributions. We translate this model into a flexible and easy-to-use quantum fault injection toolkit, which is part of our contribution and will be disclosed as open-source code~\cite{asuqa_repo}.

Our extensive analysis is based on over \textit{400 million} faults injections and considers both the repetition and XXZZ surface codes implemented in various fashions. 
We detail how physical level faults spread up to the higher-abstraction logical layer and correlate the code performance with the surface code distance, the number of physical qubits affected, and the architectural interconnection pattern of qubits. Overall, we consider more than 12 different configurations.

We show that surface code performance degrades significantly in the presence of radiation-induced transient faults, reaching logical error rates of up to $54\%$ and that a single particle interaction that spreads to neighbour qubits has an effect on the code output which is worse than corrupting half of the available qubits. Our analysis highlights that, with a constant number of physical qubits, the bit-flip repetition code is up to 10\% more effective against radiation than XXZZ. Moreover, by properly selecting the underlying hardware topology we can improve the radiation fault correction by up to 7\% in the repetition code and 9\% in the XXZZ code.


While several works have evaluated radiation's effect in quantum devices~\cite{Barends2011, radiation2011, LossMechanisms2018, nature_rad, Wilen2021, Cardani2021, Martinis2021, Chen2021, Acharya2023} and propose some physical implementation improvements~\cite{Iaia2022, mcewen2024resisting, Acharya2023, li2024direct, Ravi2023, Sivak2023}, to the best of our knowledge this is the first paper to target the effectiveness of surface codes in correcting radiation-induced events. 

The remainder of the paper is organised as follows. 
Section~\ref{background_and_related_works} sums up domain specific knowledge of quantum noise modelling and surface code formalism, whilst Section~\ref{system_simulation_model} describes the implementation of both the intrinsic noise and radiation-induced fault models. 
In Section~\ref{exploration_of_design_space}, we define the surface code classes that have been tested, to later present an analysis of the collected data in Section~\ref{results}.
We 
conclude the work by summarising the main results and digressing on the future investigation paths 
in Section~\ref{conclusions_and_future_works}.

\section{Background and related works}
\label{background_and_related_works}
In this Section we 
summarize the background and formalism on quantum computing, reliability, and surface codes necessary to follow along with the remainder of the paper. 

\subsection{Quantum Computing}
Quantum computing encodes information in controllable two-level quantum mechanical systems, generally referred to as qubits. Amongst the physical systems employed 
for implementing quantum computers, we find superconducting qubits implemented with Josephson junctions, trapped ions, neutral atoms, photonic devices and many more~\cite{Wendin2017}. 
\textbf{This paper focuses on the superconducting  qubit}, as it is the most adopted and promising technology currently available and the only one for which the radiation impact has been extensively characterized~\cite{Barends2011, radiation2011, LossMechanisms2018, nature_rad, Wilen2021, Cardani2021, Martinis2021, Chen2021, Acharya2023}.

Given their quantum nature, qubits exhibit special properties that can be leveraged to gain a computational advantage over classical algorithms, namely \textit{superposition} and \textit{entanglement}.
Superposition ties into the probabilistic nature of quantum information, as a qubit can simultaneously exist in the two basis states at once with a given probability amplitude. 

\begin{equation}
    \ket{\Psi} = \alpha \ket{0} + \beta \ket{1}, \quad\quad \alpha^2 + \beta^2 = 1
\end{equation}
Entanglement describes the ability of two qubits to share a non-classical correlation that acts on the information stored in two or more qubits.
This 
lets us infer information about the state of multiple qubits by only measuring a subset of them.
Measuring a qubit inherently destroys the quantum information it stores, projecting it to a classical bit.

\begin{figure*}[!th]
    \centering
    \includegraphics[width=\linewidth]{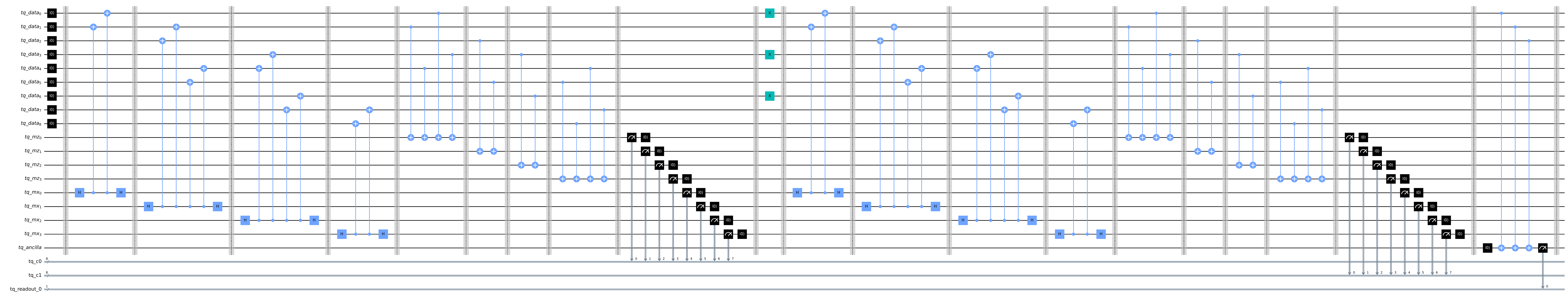}
    \caption{\textbf{Quantum circuit representation}. The Distance-(3,3) XXZZ surface code.}
    \label{xxzz_code_circuit}
\vspace{-10pt}
\end{figure*}

\begin{figure}[!th]
    \centering
    \includegraphics[width=\linewidth]{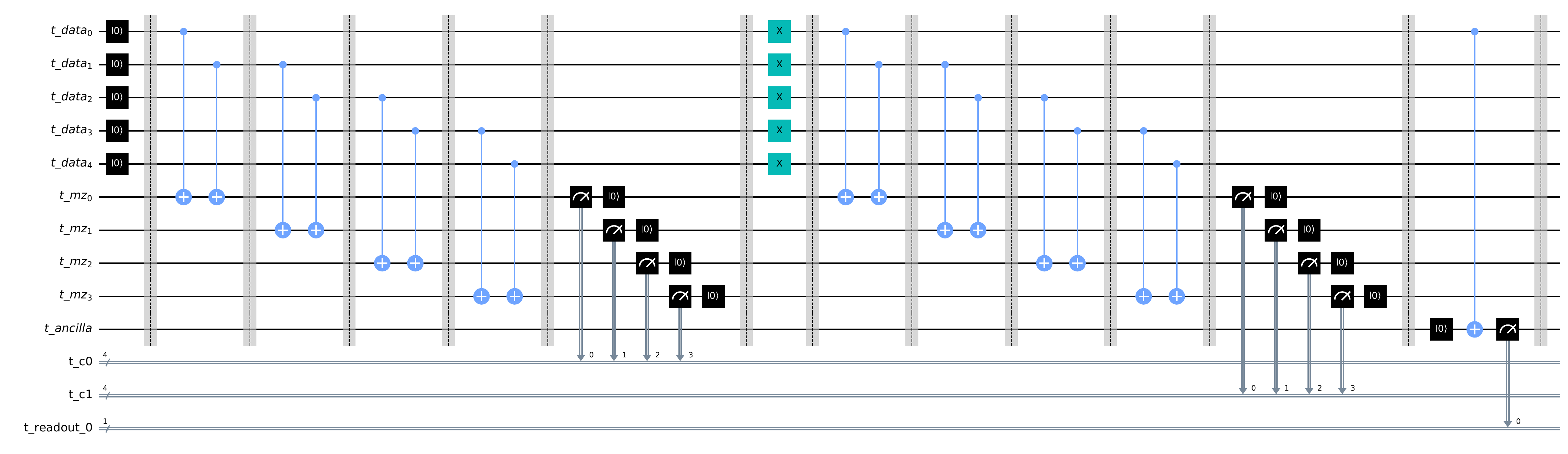}
    \caption{\textbf{Quantum circuit representation}. The Distance-5 bit-flip protected repetition code.}
    \label{repetition_code_circuit}
    \vspace{-10pt}
\end{figure}

Sequences of qubits, often referred to as quantum registers, can be acted upon through the application of 
unitary operators, or \textit{quantum gates}.
Algorithms described through this gate-based formalism take the name of quantum circuits, 
as shown in Figures~\ref{xxzz_code_circuit} and~\ref{repetition_code_circuit}.
Given the probabilistic nature of qubits, quantum computers' results are expressed as probability distributions of observed bit-strings over multiple (read: \textit{thousands}) of identical circuit executions.

Quantum computers have a specific hardware topology, i.e., possible inter-qubits connectivity. The process of transpilation maps the logical circuit in the underlying topology, adapting the logical connections with the available ones.

\subsection{Decoherence, thermal noise and measurement errors}
\label{intrinsic_noise_teoretical}
Each qubit implementation has intrinsic reliability issues. Superconducting quantum computers, which are the focus of our paper, to sustain the superconducting regime of the Josephson junction, are to be operated at temperatures in the \textit{mK} regime, and need to be completely isolated from the environment to correctly retain their quantum properties, something that can hardly be reached due to engineering limits~\cite{Unruh1995}.
These requirements stand in stark contrast with device operability, as interacting with it will necessarily introduce unwanted noise.
This gives rise to two main time metrics used to describe noise in quantum computers: \textit{spin-lattice coherence time ($T_1$)} and \textit{spin-relaxation time ($T_2$)} \cite{Preskill_2018}.
$T_1$ rules the time dependent exponentially decaying probability distribution over which a qubit can retain quantum information before collapsing to the ground state. 
$T_2$ defines the time dependent exponential decay during which a qubit will transition from a superposition state to a classical mixture of basis states.
These metrics are used to model the decoherence of a two-level quantum system.
Intrinsic noise also encompasses the accuracy of the controllers that initialise and read the state of qubits, with \textit{state preparation and measurement} errors.

Pauli-based noise models are universally employed to simulate intrinsic noise properties of quantum computers through unitary operators, since they have been validated to be sufficiently accurate~\cite{Georgopoulos2021}.
As detailed in Section~\ref{sub_noise_model}, this is the model of intrinsic noise we adopt.


\subsection{Radiation-induced faults}
\label{back_radiation}
Particle impacts, which naturally occur as a byproduct of cosmic ray decay, 
have been shown to induce faults in quantum devices~\cite{LossMechanisms2018, nature_rad, Wilen2021, Cardani2021, Oliveira2023neutrons, Martinis2021, Chen2021, Acharya2023, harrington2024synchronous}.
These unavoidable stochastic natural alter the state of qubit(s) by forcing them into decoherence for long periods of time.
The fault mechanism involves the generation of electron-hole pairs in the silicon substrate of the quantum chip, which in turn break cooper pairs in the Josephson junction forming quasiparticles, that rapidly give rise to long-lasting phonons, responsible for spreading the energy across the lattice of the quantum computer's substrate and interconnections~\cite{LossMechanisms2018, nature_rad, Wilen2021}.

The fault generation in qubits and the consequent impact on the logic value are significantly different from classical CMOS technology. A transistor state is momentarily reversed if (and only if) the deposited charge is higher than a threshold value (critical charge) that depends on the technology~\cite{Baumann2005}. Most impinging particles (such as muons, gamma rays, or low energy neutrons) do not deposit sufficient charge to trigger the fault in a transistor~\cite{Baumann2005}.
In a qubit, even a single Cooper pair break is sufficient to disturb the quantum equilibrium, thus modifying the logic status~\cite{nature_rad, Wilen2021}. Since the energy required to break a Cooper pair is in the order of milli-electronvolts (meV), even very light interactions are sufficient to induce a fault~\cite{Oliveira2023neutrons}, exacerbating the probability of observing a radiation-induced fault in quantum devices.
Field experiments performed by Google AI on a 25 qubits array showed radiation-induced faults every tens of seconds~\cite{mcewen2022resolving}. The reported error rate is several orders of magnitude higher than the one of modern CMOS technology. As a reference, the whole Titan supercomputer (composed of 14,000 nodes) has an error rate in the order of one error every few hours~\cite{hpca2015}.

Radiation is known to corrupt various devices, in the sense that one single impinging particle interacts with various transistors or qubits. With the shrink of CMOS transistors feature size, the probability of having multiple bit upsets is increasing~\cite{Baumann2005}, being over 10\% in 28nm or newer technologies~\cite{selse2014}, undermining Error Correcting Code (ECC) efficacy. Because a significantly lower charge is sufficient to affect the qubit and since the charge spread isotropically in the silicon substrate from the particle impact, the probability of having multiple qubits affected by a single particle is exacerbated. In the experiment performed by Google AI each radiation strike modified the state of the vast majority of the qubits \cite{mcewen2022resolving}.

The radiation impact is transient in nature, as the energy absorbed by the (quantum) chip is gradually dissipated by recombination soon after the impact and by diffusion later~\cite{Baumann2005}. Unfortunately, while for CMOS devices the fault duration is in the order of nanoseconds~\cite{Baumann2005}, in qubits the observed effects last from $25$ \textit{ms} to upwards of $100$ \textit{seconds}~\cite{Oliveira2023neutrons}. This time scale is orders of magnitude greater than the execution time of a single quantum circuit, making repetition ineffective.

A particle impact, then, induces a correlated stochastic event that affects multiple neighbouring qubits and lasts for upwards of seconds.
We use these facts and the physical laws that rule energy deposition and distribution in silicon to model the radiation-induced transient fault detailed in Section~\ref{sub_fault_model}.

Up to now, most of the solutions proposed to tackle radiation issues are relative to alternative qubit hardware implementations, by employing charge trapping wells and better isolating the Josephson junctions~\cite{Iaia2022, mcewen2024resisting, Acharya2023, li2024direct}, to the usage on-chip hardware solutions for simultaneous error correction that target only frequent and small uncorrelated errors~\cite{Ravi2023, Sivak2023}, or to employing deep underground facilities
~\cite{loer2024abatement}.
These solutions, albeit promising, inevitably incur in an increase production costs due to additional engineering and fabrication processes.
This may be bound to follow a similar trend to what happened in the case of transistors, as it is preferable to find flexible software solutions, rather than expensive hardware ones.

\subsection{Surface codes}
Quantum computers can be error-corrected, through the usage of dedicated codes.
The main caveat 
is that, due to the no-cloning theorem \cite{Wootters1982nocloning}, one cannot copy an arbitrary quantum state.
As such, error correction in quantum systems is not as straightforward as adapting classical correction codes to quantum circuits.
The idea is to use a portion of \textit{data} qubits to encode the information, and use a portion of \textit{observer} qubits to detect error syndromes.
At the foundation of quantum error correction (QEC) we find the stabiliser formalism.
A stabiliser is the combination of CNOT gates applied to some \textit{data} qubits, storing the information we wish to preserve, and \textit{observer} qubits, that give us information about an error syndrome in a given basis.
Commonly used syndrome bases are the Z-basis for bit-flips and the X-basis for phase flips, although it must be noted that the set of possible syndromes is infinite, as it is possible to define an infinite number of arbitrary unitary operators on a qubit.
As such, the error correction capabilities of surface codes are always going to be limited by the expressiveness of the syndrome set chosen.
\begin{equation}
    X = \begin{bmatrix}
        0 & 1 \\
        1 & 0 
        \end{bmatrix}
    , \quad X\ket{0} = \ket{1}
\end{equation}
\begin{equation}
    Z = \begin{bmatrix}
        1 & 0 \\
        0 & -1 
        \end{bmatrix}
    , \quad Z\ket{1} = -\ket{1}
\end{equation}
By measuring and comparing observer qubits through 
a \textit{stabiliser measurement}, it is possible to infer the presence of error syndromes in the data qubits, and consequently applying the reverse operators 
to undo the error \cite{JavadiAbhari2017, Andersen2020, Krinner2022}.

Surface codes spread information over multiple physical qubits, where data qubits share  sentinel qubits following a specific pattern.
The first such implementation of a surface code was presented on a bidimensional mesh with periodic boundaries (i.e. a torus)\cite{Kitaev2003}.
The idea of this 
approach is to spatially relate all of the stabiliser measurements that detect an error over a decoding step.
This is done by matching the error syndromes closest to each other, forming loops that can be shrunk down to a point and then applying the corresponding syndrome correcting operators. 

The technique generally employed in the literature to perform error syndrome decoding consists in using minimum weight perfect matching (MWPM) algorithms \cite{Brown2022, Marton2023, Vittal2023, Sundaresan2023, higgott2023sparse}.
Other approaches exist, such as ones based on tensor-networks \cite{Bravyi2014}, belief propagation \cite{Old2023}, union-find \cite{Delfosse2021}, or machine learning \cite{Varsamopoulos2018}.
However, given that MWPM offers the better trade-off between high accuracy and low time-to-solution, testing other decoders it outside of the scope of this article.

\section{Noise and Fault Model Formalisation}
\label{system_simulation_model}
This Section describes how the previously introduced intrinsic and radiation noise have been modelled to perform efficient and accurate simulations of the quantum computer's behaviour.

\subsection{Intrinsic noise model}
\label{sub_noise_model}
A superconducting device's ability to retain information varies over time, and depends on the accuracy associated with performing each quantum gate operation.
As such, noise models are necessary in order to accurately simulate the behaviour of real quantum computers.

Following the common practices found in the literature, we decided to use a depolarisation error model based on \textit{unitary} Pauli operators, adapted from \cite{Georgopoulos2021}.
The uncorrelated nature of this noise model's faults follows the definitions of intrinsic noise provided in the literature \cite{Preskill_2018}, and surface codes are built and optimised against this kind of depolarisation noise.
The model we use is parameterised over a \textit{physical error rate} $p$, and acts by appending an X, Y or Z operator after each gate operation $\mathcal{O}$ with probability $\frac{p}{3}$, thus producing uncorrelated errors in time and space:
\begin{equation}
    \label{bitphase_flip_model}
    \mathcal{O}\ket{\psi} \xrightarrow{}\mathcal{E} \mathcal{O}\ket{\psi} \text{with} \ \mathcal{E} \doteq \sqrt{\scriptstyle{1-p}} \ \mathbb{I} + \sqrt{\scriptstyle{p/3}}\left(X+Y+Z \right).
\end{equation}
When performing two-qubit gate operations, we append after each of them an error gate obtained as the tensor product of two independent $\mathcal{E}$ noise operators: $\mathcal{E}_2 \doteq \mathcal{E} \otimes \mathcal{E}$.

Such an error model is frequently used in literature to benchmark the performance of surface codes, and is inherently defined as one of the main examples of a \textit{nice} error basis.
Given that surface codes have relatively low circuit depth, 
their execution time is orders of magnitude lower than the characteristic T1 and T2 times of modern superconducting quantum computers \cite{Somoroff2023, ganjam2023surpassing}.
As such, the state coherence difference between the first and last gate operation in the circuit can be approximated as being constant without incurring a loss in simulation fidelity.

\subsection{Radiation-Induced fault model}
\label{sub_fault_model}
As detailed in Section~\ref{back_radiation}, the impact of radiation breaks the quantum equilibrium, causing a loss of coherence.
To model the impact of radiation in the logic state of a qubit, we append a \textit{non-unitary} reset operation to each quantum gate acting on that qubit with probability $p_{q_i}$, with $i$ being the qubit's index.
The energy deposited by the impinging particle, and thus the probability of applying the reset operation on a qubit, depend on the distance from the point of impact and decays over time.
As such, the fault event we model evolves both in the spatial and temporal domains from the \textit{root impact point}, i.e. the qubit from which the fault spreads.

In the \textbf{time domain}, since the deposited charge in Silicon recombines and diffuses~\cite{Baumann2005}, the fault event evolves as a decaying exponential~\cite{yelton2024modeling, harrington2024synchronous, Baumann2005} that spikes at the root impact point and wears off to zero as time goes on.
In Equation~\ref{temporal_decay} we detail the temporal decay function $T(t)$ over continuous time $t \in [0, 1]$, that outputs the probability of generating quasiparticles in the Silicon substrate:
\begin{equation}
    \label{temporal_decay}
    T(t) = e^{-\gamma t}, \quad\quad \gamma = 10, \quad\quad t \in [0, 1] \ .
\end{equation}

\begin{figure}[!t]
    \centering
    \includegraphics[width=1\linewidth]{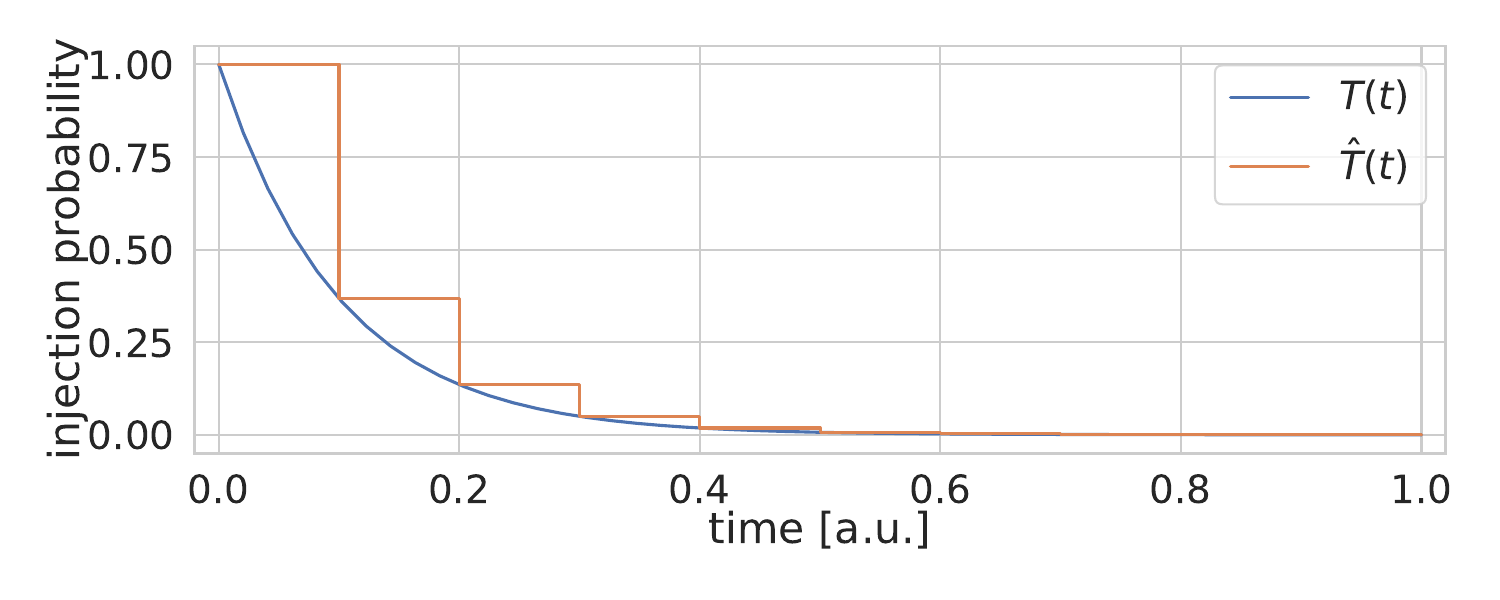}
    \caption{\textbf{Intensity of the radiation-induced fault according to time}. The temporal  decay function $T(t)$ parameterised over time and the approximated function $\hat{T}(t)$ over $n_s=10$ samples. The time axis is in arbitrary units.}
    \label{temporal_damping_evolution}
    \vspace{-10pt}
\end{figure}

The factor $\gamma = 10$ defines the exponentially decaying presence of quasiparticles in the Silicon substrate, following the experimental rates highlighted in the literature \cite{nature_rad, mcewen2022resolving, Chen2021}.
The estimated time required to execute a shot of the surface code on a real quantum computer ranges, on average, from $\sim 14$ $\mu s$ to $\sim 125$ $\mu s$ \cite{Acharya2023, lin2024codesign, xiao2024exact}.
Transient radiation-induced faults can last for upwards of $100$ seconds \cite{mcewen2022resolving, Oliveira2023neutrons}.
Given this order of magnitude difference, we can approximate the time evolution of $T(t)$ with a step function $\hat{T}(t)$, by sampling the temporal decay $T(t)$ over $n_s$ number of samples, as shown in Figure \ref{temporal_damping_evolution}.
We selected $n_s = 10$, meaning that the function $T(t)$ has been sampled over 10 equidistant points in time, granting a reasonable trade-off between accuracy and performance.
Increasing the number of samples comes at the expense of computational overhead.

\begin{figure}[!t]
    \centering
    \includegraphics[width=0.9\linewidth]{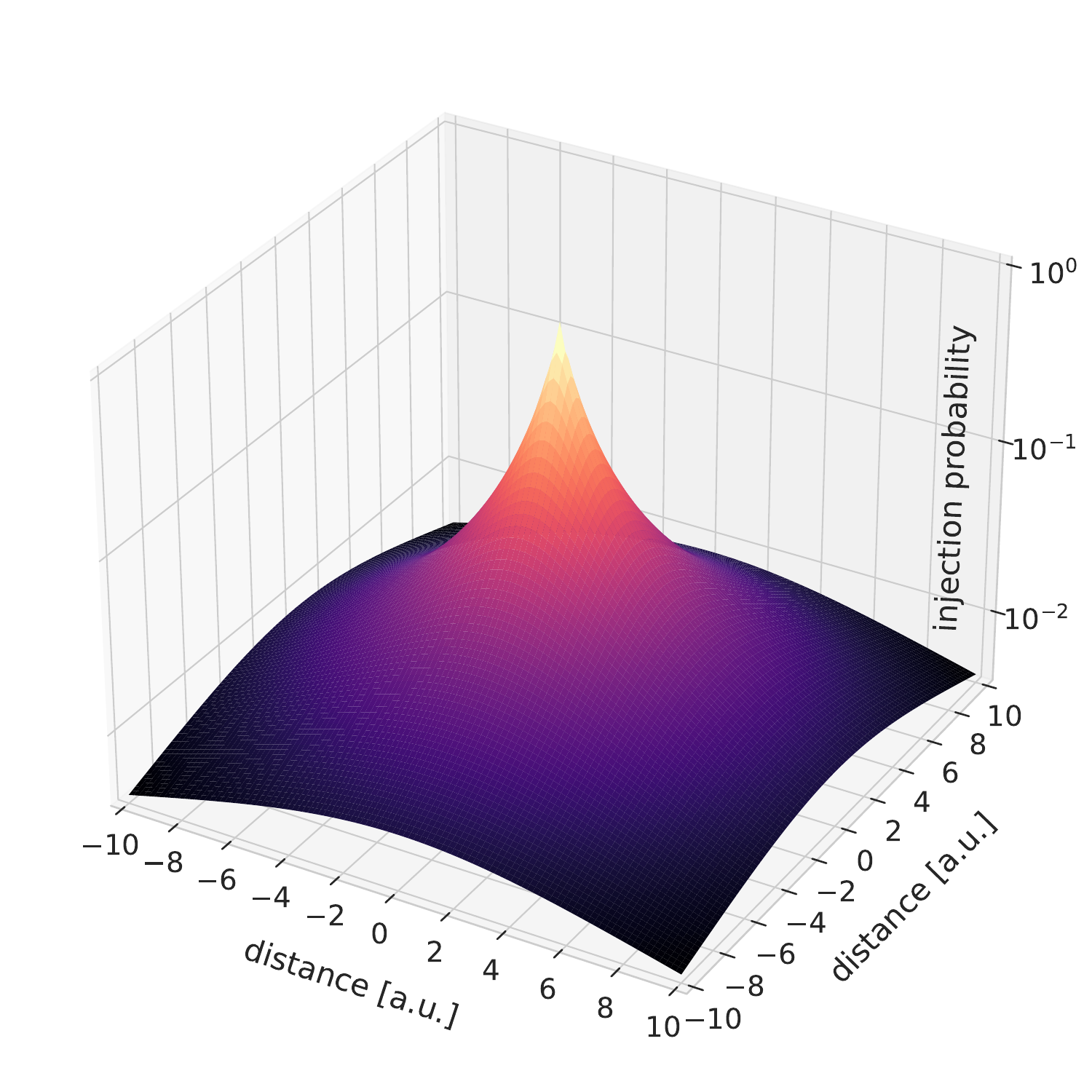}
    \caption{\textbf{Intensity of the radiation-induced fault according to distance}. The spatial decay function $S(d)$ parameterised by the distance from the root impact point at coordinate $(0,0)$, with a peak of $100\%$. Both distance axes are in arbitrary units.}
    \label{spatial_damping_evolution}
    \vspace{-10pt}
\end{figure}

In the \textbf{spatial domain}, the deposited charge spreads from the root impact point throughout the quantum chip, diminishing in intensity the further a qubit lies from the impact point~\cite{Wilen2021}.
By following the qubit interconnections in the quantum computer's architecture, we devise an undirected graph, called architecture graph.
We consider a fixed weight on each edge of $n=1$.
This behaviour approximates the electron hole pair distributions induced by particle impacts in Silicon over a normalised integer distance $d$ \cite{Wilen2021}.
We then define a spatial damping function $S$, parameterised by the minimum distance between two qubits in the architecture graph:
\begin{equation}
    S(d) = \frac{n^2}{(d+n)^2}, \quad n = 1 \ .
\end{equation}
To parameterise the application of faults on a qubit, the root injection probability sampled from $T$ is thus multiplied by the output of $S$.
The product of the temporal and spatial domain fault evolution functions is collectively defined as $F$, the transient error decay function:
\begin{equation}
    F(t, d) = T(t) S(d) \ .
\end{equation}
The fault probability $p_{q_i}$, obtained from sampling $F(t, d)$, is computed on a per-qubit basis and it is used to parameterise the application of a reset operator after each gate applied to that qubit.
An example of the spatial evolution of the root injection probability at time $t=0$ is presented in Figure \ref{spatial_damping_evolution}.

\section{Exploration of Design Space}
\label{exploration_of_design_space}
For our anayslis we considered two of the most widely employed surface codes: the repetition 
and the XXZZ surface code.  
The library used to generate the parameterised correction codes and decoding their output with the MWPM algorithm is courtesy of the Qiskit Topological Codes project~\cite{qtcodes}.
While we present an extensive evaluation on two cornerstone codes, the methodology we employ and the analysis we propose are not tightened to a specific code or implementation and can be easily adapted to future QEC codes, as they become available.

\begin{figure*}[!ht]
\centering
\begin{subfigure}{.49\textwidth}
  \centering
  \includegraphics[width=1.00\linewidth]{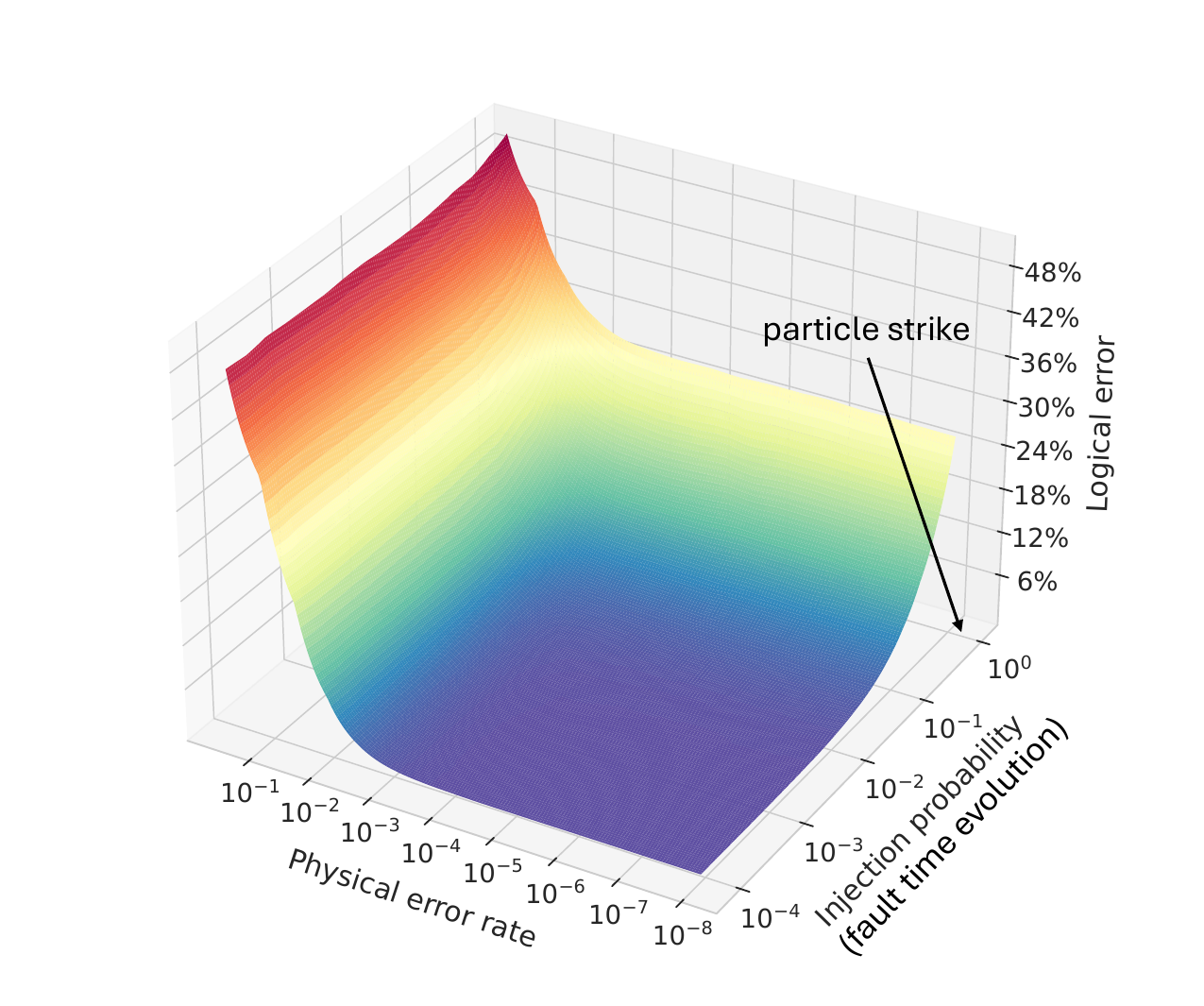}
  \caption{Distance-(5,1) bit-flip repetition code.}
  \label{surface_repetition}
\end{subfigure}%
\begin{subfigure}{.49\textwidth}
  \centering
  \includegraphics[width=1.00\linewidth]{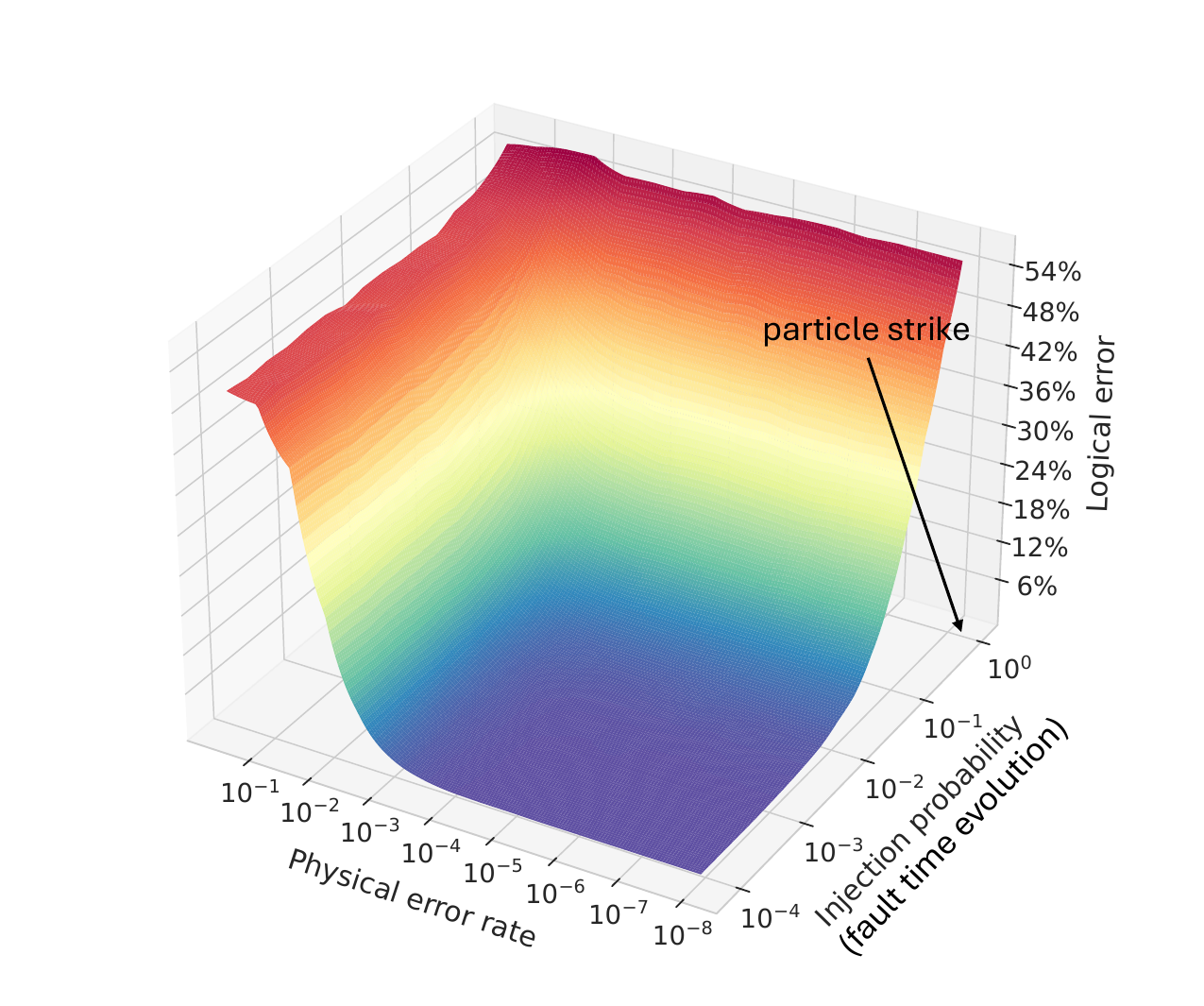}
  \caption{Distance-(3,3) XXZZ code.}
  \label{surface_xxzz}
\end{subfigure}
\caption{\textbf{Logical error landscape}. The two plots show the combined impact of intrinsic device noise and radiation-induced faults. On the bottom right axis we see the time evolution of the radiation fault parameterised by the root injection probability (100\% at time of impact), whilst on the bottom left axis we see the physical error rate, i.e., the intrinsic noise model.}
\label{surface_plots}
\vspace{-10pt}
\end{figure*}

\subsection{Repetition code}
The quantum repetition code employs \textit{n} data qubits to encode the state of one single logical qubit, by repeating information in what is called a \textit{GHZ state}, and additional \textit{n} qubits to perform stabiliser measurements, necessary to extract error syndromes.
The overall distance of the code is defined as $d=(d_Z, d_X)$, with either $d_Z=1$ for phase-flip protection codes and $d_X=1$ for bit-flip protection codes.
\begin{equation}
    \ket{0} \xrightarrow{} \ket{\psi_0}^{\otimes n}, \quad 
    \ket{1} \xrightarrow{} \ket{\psi_1}^{\otimes n}
\end{equation}
This code can either offer protection from bit-flips or phase-flips, according to the basis chosen for the GHZ state: by using the Z-basis, we have bit-flip protection, whilst by choosing the X-basis we have phase-flip protection.
The repetition code can detect only the corresponding encoding basis error on up to $\left \lfloor{(n-1)/2}\right \rfloor $ qubits, so long as those error events are uncorrelated (ndr, radiation events are correlated). 
The total number of qubits required to encode a repetition code is $q_{rep} = 2n$, with $n=\max(d_Z, d_X)$ being odd, and either $d_Z$ or $d_X$ being equal to $1$.
This code is one of the few that have been extensively tested on superconducting quantum computers \cite{Acharya2023, Chen2021, Wootton2020}.
As an example, the circuit structure for the distance-$(5, 1)$ quantum circuit bit-flip protected repetition code, which uses $10$ qubits, is shown in Figure \ref{repetition_code_circuit}.
The circuit pattern contains a first stabilisation component, represented by the chain of nearest neighbour CNOTs controlled by the data qubits and targeting the stabilisation qubits, followed by a round of syndrome measurements.
At the centre of the quantum circuit, in green, we find the replicated application of a logical operation (an X gate) to all the logical qubits, followed by a second round of syndrome measurement.
The code raw output is extracted by applying an ancilla readout.

\subsection{XXZZ code}
\label{xxzz_code_description}
The XXZZ surface code is a rotated surface code generated by XXZZ and ZZXX Pauli strings, clockwise associated with the vertices of each face in a two-dimensional mesh, with one qubit in each vertex.
It is virtually identical to the XZZX code, only varying in terms of Pauli strings generators for the stabiliser plaquettes \cite{BonillaAtaides2021}.
It is an adaptation of the toric code with non-periodic boundaries \cite{Kitaev2003}.
The mesh is defined over two odd integers $d_Z$ and $d_X$, one for the Z-error stabilisers (i.e. bit flips) and one for the X-error stabilisers (i.e. phase flip).
The overall distance of the code is defined as $d=(d_Z, d_X)$, with the total number of qubits required to encode these circuits being $q_{XXZZ} = 2 d_Z d_X$.
This code has been tested on superconducting quantum computers, letting researchers achieve logical error rates lower than those of the physical qubits they are encoded with \cite{xiao2024exact, forlivesi2023logical}.
As an example of the code's structure, specifically for the distance-$(3,3)$ XXZZ surface code, is shown in Figure \ref{xxzz_code_circuit}.
A total of $n=d_Z d_X$ qubits are used to encode the data, $m=\left \lfloor{\frac{(d_Z d_X) - 1}{n}}\right \rfloor $ qubits are used for measuring Z-basis errors, $m$ qubits are used to detect X-basis errors and one final ancilla qubit is used to perform the raw code readout.

\subsection{Simulation parameters}
\label{simulation_parameters}
To provide realistic injection data, we model the quantum computer's intrinsic noise as a depolarising channel, as detailed in Section \ref{sub_noise_model}.
This intrinsic noise model is parameterised with probability $p=1\%$\cite{higgott2023sparse}, unless otherwise noted. In absence of radiation-induced events all the tested configurations do not present output errors.

The approximated temporal damping function $\hat{T}(t)$ has been sampled over $n_s = 10$ equidistant points in time.

All simulations refer to a $5\times 6$ bidimensional lattice as the architecture graph, except for the architectural analysis in Section \ref{architectural_analysis}.

For both of the analysed surface code classes, we initialise each data qubit to $\ket{0}$ and encode a logical X-gate operation.
The surface code's circuit is repeated over multiple execution shots, over a simulated radiation event that lasts for $100$ \textit{ms}.
The expected logical output post-decoding of the surface code is a logical $\ket{1}$ state, as detailed in the example circuit diagrams shown in Figures \ref{repetition_code_circuit} and \ref{xxzz_code_circuit}.
Measurements are decoded through the MWPM algorithm and a logical error is detected whenever the output of the decoder is a logical $\ket{0}$ state.
The logical error rate is computed as the number of shots that are decoded as a $\ket{0}$ state divided by the total number of shots.

\section{Results}
\label{results}
We performed a total of four different analyses over two quantum error-correcting code classes.
When referring to \textit{injection data}, we consider the whole time and space evolution of a given fault, unless otherwise noted.

\begin{figure*}[!ht]
\centering
\begin{subfigure}{.49\textwidth}
  \centering
  \includegraphics[width=1.00\linewidth]{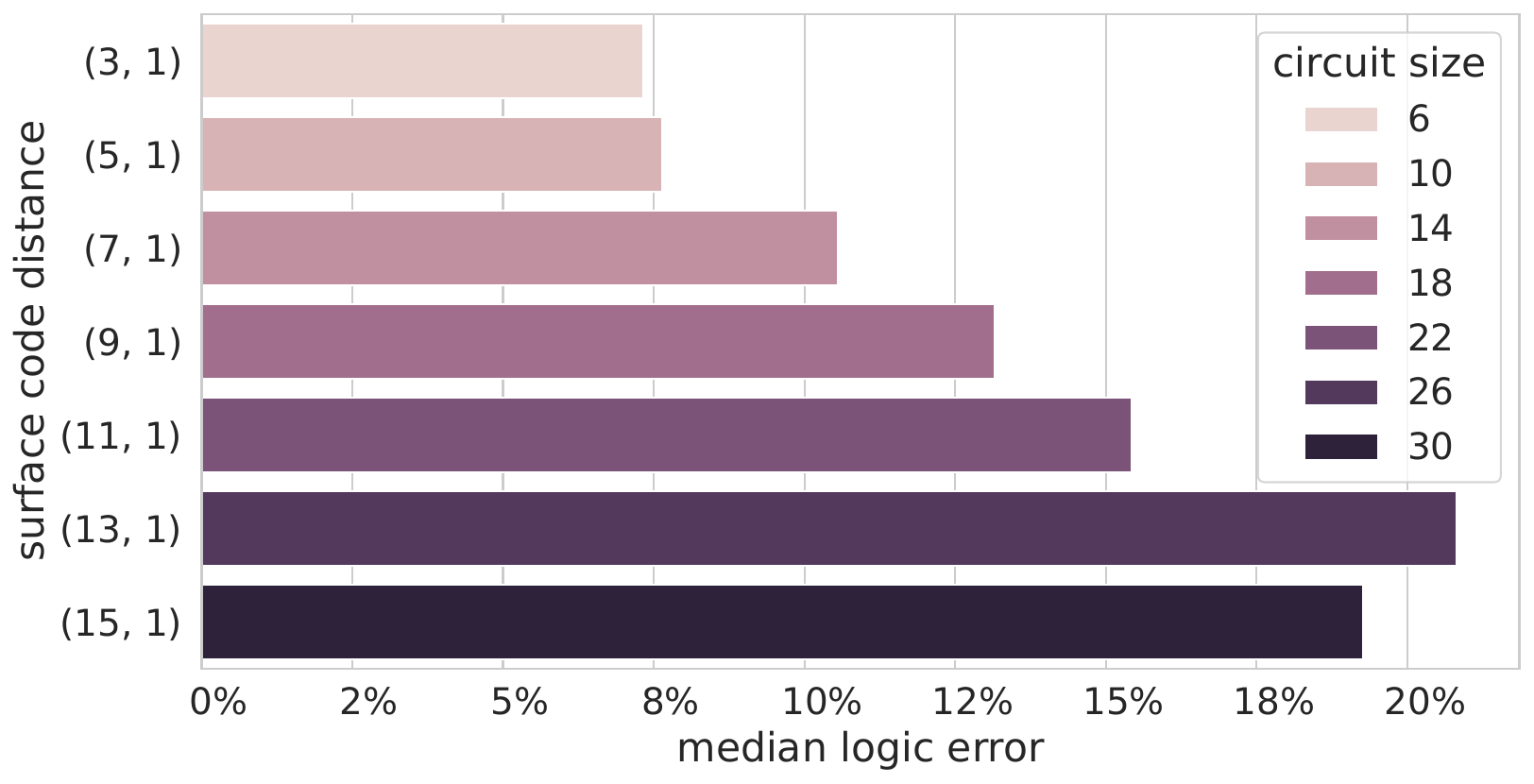}
  \caption{Bit-flip repetition code.}
\end{subfigure}%
\begin{subfigure}{.49\textwidth}
  \centering
  \includegraphics[width=1.00\linewidth]{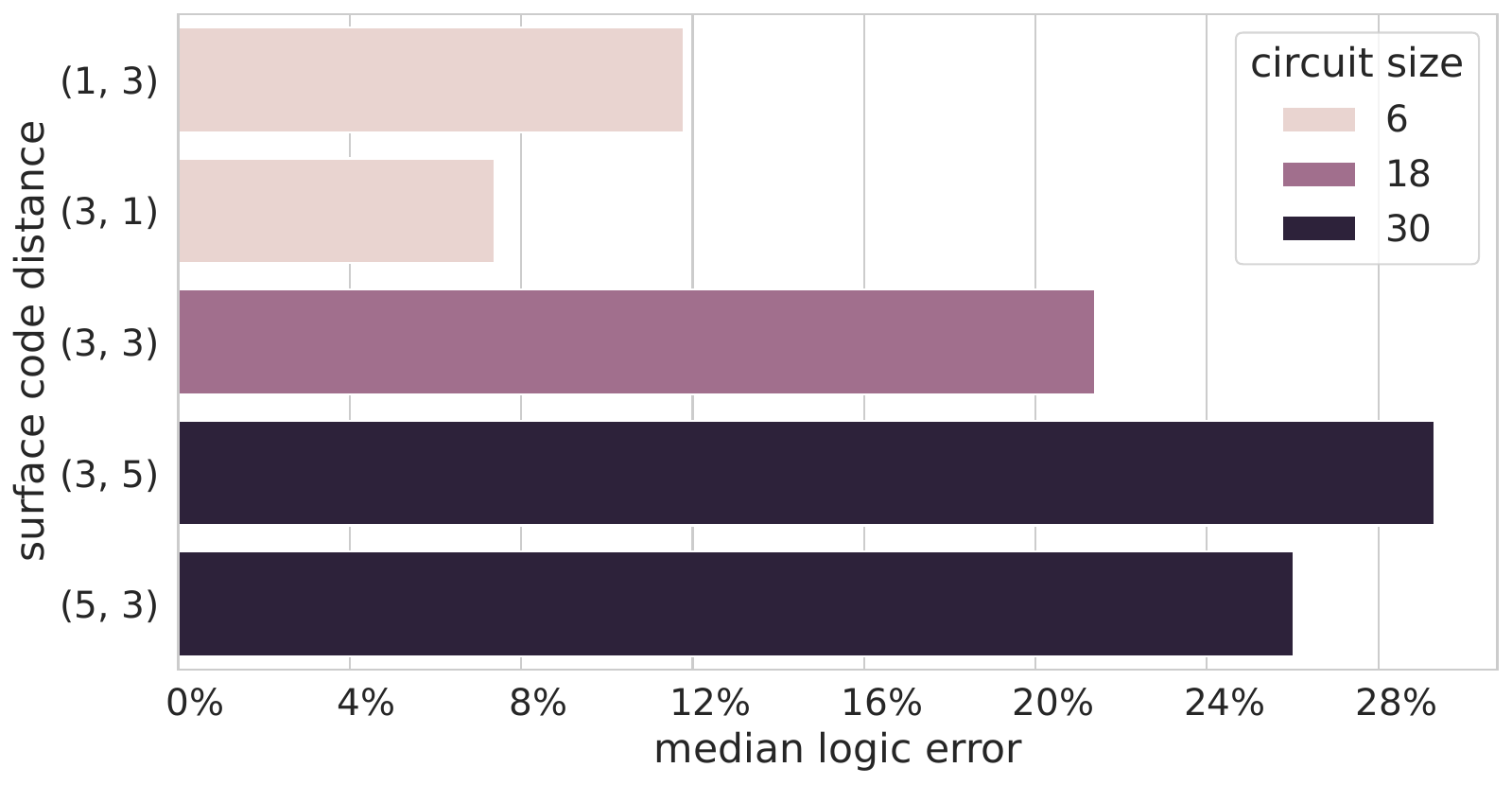}
  \caption{XXZZ code.}
\end{subfigure}
\caption{\textbf{Logical error criticality by code distance}. The effect of a single erasure error (reset) at time $t=0$ that does not spread to neighbouring qubits, for different surface code sizes. The hue represents the number of physical qubits in the surface code's circuit.}
\label{minimum_inj_qubits}
\vspace{-10pt}
\end{figure*}

\subsection{Noise vs. radiation-induced faults analysis}

Surface codes, in their various implementations, have been designed to correct intrinsic noise. 
With this first analysis, to answer the research question whether surface codes are effective in correcting radiation-induced faults, we evaluate: (1) how intrinsic noise and radiation-induced faults interoperate and (2) if the two events show interference patterns that influence the output logical error of the surface code.
To do this, we correlate the logical error rate of surface codes with respect to radiation-induced transient faults and the intrinsic noise model of a quantum computer. 
We report data on the distance-(5,1) repetition code and the distance-(3,3) XXZZ code.
Both surface codes are transpiled over a square lattice architectural layout.
Specifically, the repetition code lattice was of size $5\times2$, whilst the one for the XXZZ code was $5\times4$. 
Data from simulations with alternative parameters have unveiled a similar behaviour, and will not be further commented on.
The root injection point has been deterministically chosen to be the qubit with index two for reproducibility reasons.
A further analysis on the choice of the injection point is discussed in Section~\ref{architectural_analysis}.

In Figure \ref{surface_plots}, we plot the post-decoding logical error as a 3D plot.
One ground axis represents the injection probability, i.e. the radiation-induced fault time evolution, which has a one-to-one mapping with time as seen in Section~\ref{sub_fault_model}: at the time of strike, the probability to modify the qubit state is 100\%, exponentially diminishing as time passes.
The second ground axis represents the noise threshold $p$ of the intrinsic noise model introduced in Section~\ref{sub_noise_model}.
We test the intrinsic noise model over a range of values for $p$, the \textit{physical error rate}, going from $10^{-8}$ up to $10^{-1}$.
The higher extreme for the physical error rate has been selected to highlight the effects of intrinsic noise on a scale similar to that of the transient error, whilst the lower extreme of $10^{-8}$ is the target error rate required to reach fault tolerance in quantum computers.
A physical error rate on the order of $10^{-3}$ well approximates the noise behaviour of current quantum devices \cite{Chen2021}.
We also consider the full range of the time evolution of the transient fault, represented as the root injection probability at a given point in time.
We then interpolate the post-decoding logical error of the surface code at each coordinate.

As shown in Figure \ref{surface_plots}, at the particle strike, when the root injection probability is close to $100\%$, the considered error correction codes show an average logical error rate $27\%$ and $50\%$, respectively.
In the case of the repetition-(5,1) code, the highest value for the logical error rate of $48\%$ is reached when maximising both the intrinsic noise model's error rate, at $10^{-1}$, and the root injection probability on the repetition qubit, at $10^0$ ($100\%$).
Similarly, the XXZZ-(3,3) code peaks at $54\%$ under the same conditions.
Both surface codes achieve a logical error rate lower than the physical error rate only when the latter is smaller than $10^{-3}$ \cite{Chen2021}.
This matches the surface code performance metrics presented in surface code simulation works \cite{higgott2023sparse}.
Crucially, when reducing the physical error rate of the simulation to a regime which is unreachable for current quantum computers, such as $10^{-8}$, we still see the detrimental effects of the radiation-induced fault, as the logical error reaches $24\%$ for the repetition-(5,5) code and $52\%$ for the XXZZ-distance(3,3) code.
This gives us a clear insight: regardless of the gate level accuracy of current or future quantum computers, radiation-induced faults will still catastrophically corrupt the outputs of error correction codes.
As such, reaching extremely low qubit error rates will not be sufficient to counteract radiation-induced fault events.

\obs{Particle impacts undermine surface code performances regardless of intrinsic qubit physical error rate.}

The interaction of intrinsic noise and radiation-induced faults only show constructive interference, amplifying the overall error rate of the quantum computer, as we notice no sudden pits on the surface.
This lets us infer that no recorded injection event has positively altered the output of the surface code.
As such, the intrinsic noise model has proven to act as a lower limit to the accuracy of the surface code.

\obs{Radiation-induced faults do not cause positive alterations of the surface code's output by reversing the effects of intrinsic noise, but rather amplify the logical error.}

\begin{figure*}[!ht]
\centering
\begin{subfigure}{0.5\linewidth}
  \centering
  \includegraphics[width=1.00\linewidth]{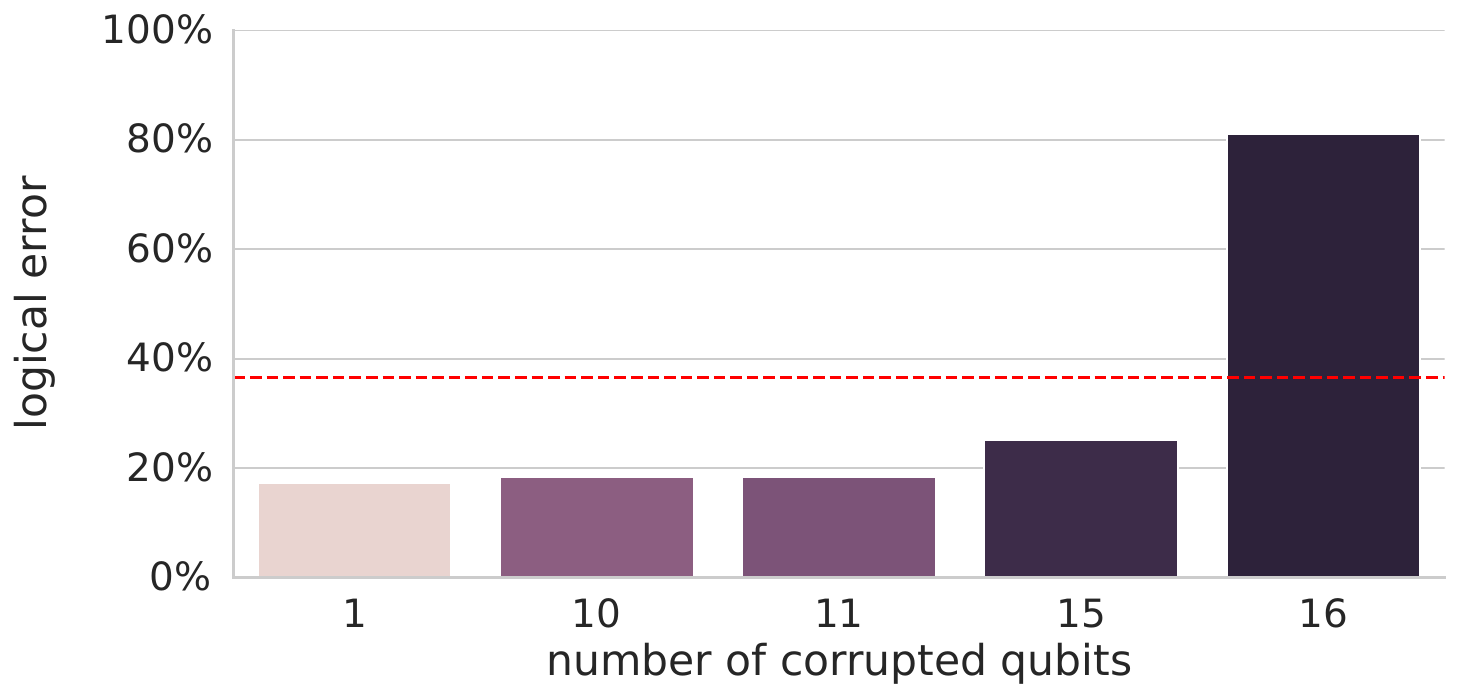}
  \caption{bit-flip distance-(15,1) repetition code.}
\end{subfigure}%
\begin{subfigure}{0.5\linewidth}
  \centering
  \includegraphics[width=1.00\linewidth]{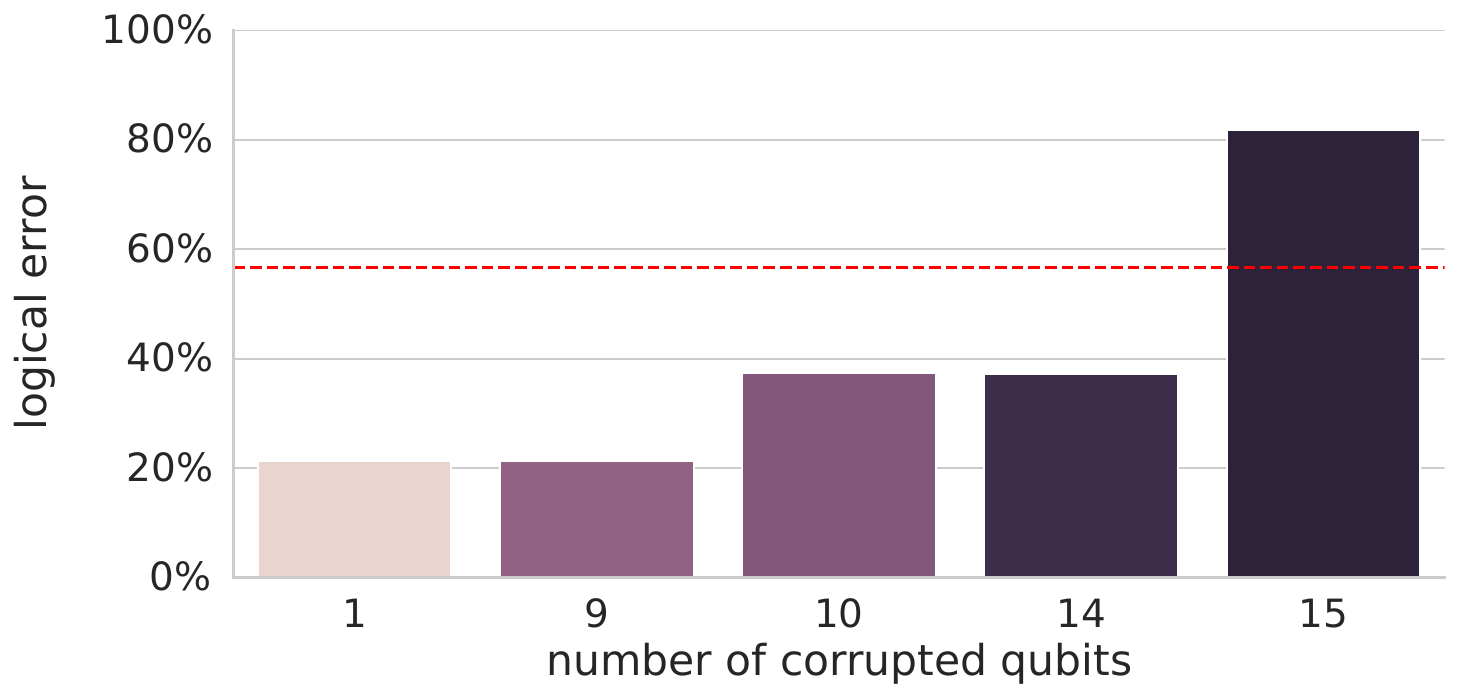}
  \caption{XXZZ distance-(3,3) code.}
\end{subfigure}
\caption{\textbf{Impact of fault spread on logical error}. Logical error caused by increasing number of corrupted qubits (reset error) compared to the logical error of a single spreading radiation-induced fault (red line). The plotted data refers only to the intensity of the fault at time $t=0$.
}
\label{bar_chart_spread_depth}
\vspace{-10pt}
\end{figure*}

\subsection{Code Distance analysis}

The repetition and XXZZ code classes are parameterised by distance, which is represented by the tuple $(d_Z, d_X)$.
$d_Z$ represents the number of qubits devoted to correct bit-flip errors, while $d_X$ represents the same statistic for phase-flip errors.
There are two research questions we aim to answer:
(1) Does a larger surface code provide better protection from radiation-induced faults?
(2) Does using both bit-flip and phase-flip protection in the XXZZ code improve the performance over the exclusively bit-flip protected repetition code? 

To answer the two research questions, we correlate the code distance with the logical error over the two considered surface code classes.
For each surface code distance, we considered one corrupted qubit, highlighting the fault's magnitude at time of impact ($t=0$).
We furthermore removed the fault's spatial expansion to neighbouring qubits, whose impact is analysed in detail in Section~\ref{spreading_vs_erasure}.
This has been done in order to highlight the moment of maximal criticality of the \textit{non-unitary} reset error, which occurs at the beginning of the event.
Each code followed the interconnection constraints of a lattice of size $5\times6$, scaled down according to the qubit requirements of each code class and distance.
We selected a subset of connected subgraphs in the lattice, then treated each subgraph as a hypernode inside of which each qubit would undergo the same fault event.
The results have then been grouped by the size of the subgraphs, extrapolating the median error across all subset sizes.

In Figure~\ref{minimum_inj_qubits} we plot the post-decoding logical error on the x-axis, and the surface code distance on the y-axis.
The bit-flip repetition code class boasts a logical error rate of $\sim 8\%$ at distance-(3,1).
For higher distance versions of the code, the logical error steadily increases, reaching a peak of $20.5 \%$ in the distance-(13,1) repetition code.
A slight logical error difference is observed in the distance-(15,1) repetition code, with a logical error of $19.5\%$, which is to be attributed to statistical noise.

The XXZZ code boasts both bit-flip and phase-flip error correction capabilities, as detailed in Section \ref{xxzz_code_description}.
In the distance-(1,3) case, we see a logical error rate of $\sim 12\%$, while in the distance-(3,1) case an error rate of $\sim 7.5 \%$ is registered instead.
When considering the distance-(3,3) case, the logical error reaches about $\sim 21\%$.
In the distance-(3,5) and distance-(5,3) codes, we see a similar behaviour as for the distance-(1,3) and the distance-(3,1) codes, only with higher logical error rates, of $\sim 29.5\%$ and $\sim 26 \%$.

\obs{Larger surface codes are more sensitive to radiation-induced faults, reaching larger logical error rates in the presence of the same fault intensity.}

This is especially highlighted in the bit-flip repetition code plot in Figure \ref{minimum_inj_qubits}, as given a single non-unitary and non-spreading erasure error, the code will generally perform worse.
This goes in contrast with the fact that under sufficiently low intrinsic noise thresholds, larger surface codes would imply lower error rates.

For like-sized surface codes, bit-flip protection stabilisers are up to $10\%$ more effective at dealing with radiation-induced errors when compared to phase-flip protection stabilisers.
\obs{bit-flip protection in surface codes is more efficient at dealing with radiation-induced faults.}
This is noticeable in the XXZZ code plot of Figure \ref{minimum_inj_qubits}, as the distance-(3,1) code and the distance-(5,3) code outperform their respective distance-(1,3) and distance-(3,5) counterparts.
This checks out as the erasure error introduced when modelling qubit corruption is a Z-basis transformation.

\subsection{Spreading fault vs. erasure fault}
\label{spreading_vs_erasure}

Particle strike events can hinder multiple qubits at once, spreading throughout the quantum chip as radiation-induced faults, a behaviour starkingly different from that of events characterising intrinsic noise.
In this analysis we answer these questions:
(1) How many simultaneous reset operations are needed to approximate the effects of a single spreading radiation-induced fault?
(2) What is the impact of a spreading radiation-induced fault when compared to a fault that does not evolve over the spatial domain?

To answer these questions, we select the connected subgraphs over a $5\times6$ lattice architectural graph, then inject all qubits in the subgroup with the same reset event, extracting the median error across all subset sizes, comparing it to the horizontal line of the radiation-induced logical error.
We highlight the surface code performance at time $t=0$, as it is the most critical moment in the evolution of the radiation-induced fault.
On this configurations, we consider the reset errors to impact only the root injection points, leaving neighbouring qubits unaffected.

\begin{figure*}[!th]
\centering
\begin{subfigure}[b]{0.52\linewidth}
    \centering
    \includegraphics[width=1.00\linewidth]{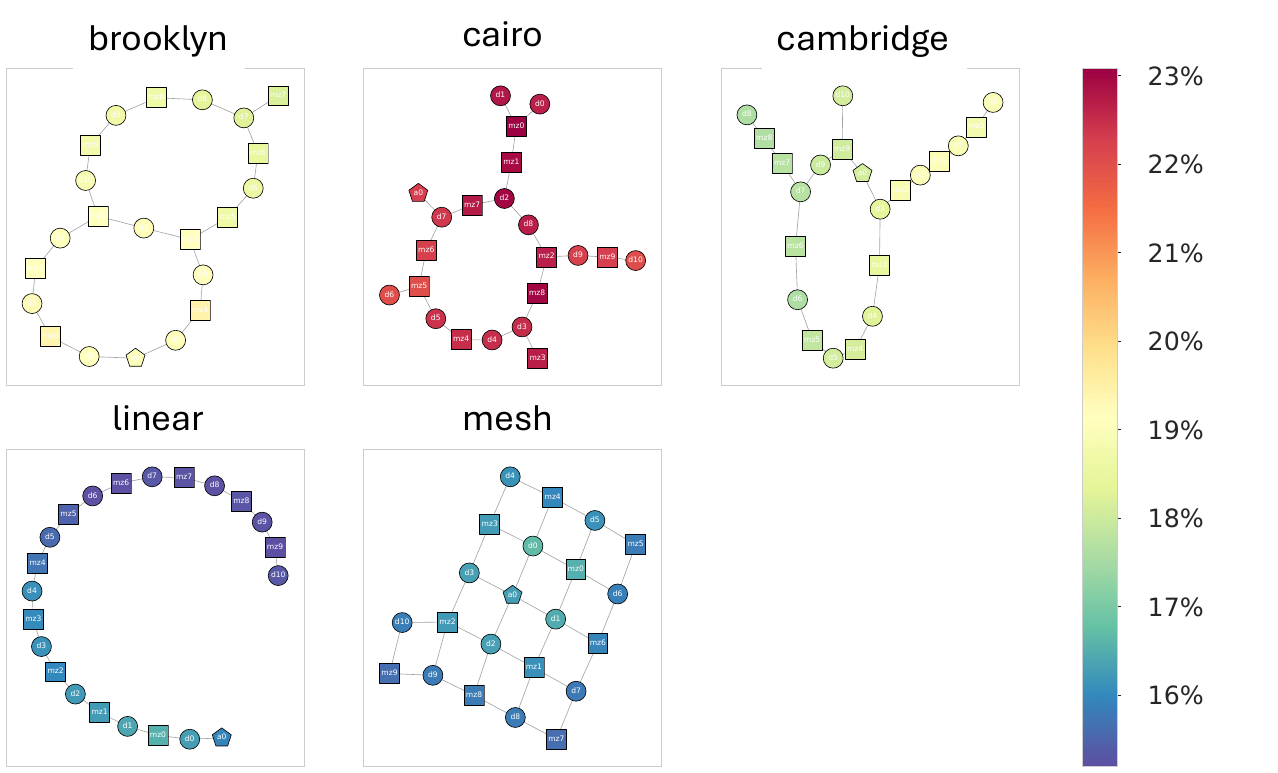}
    \caption{Distance-(11,1) bit-flip repetition repetition code.}
    \label{arch_repetition}
\end{subfigure}%
\begin{subfigure}[b]{0.52\linewidth}
    \centering
    \includegraphics[width=1.00\linewidth]{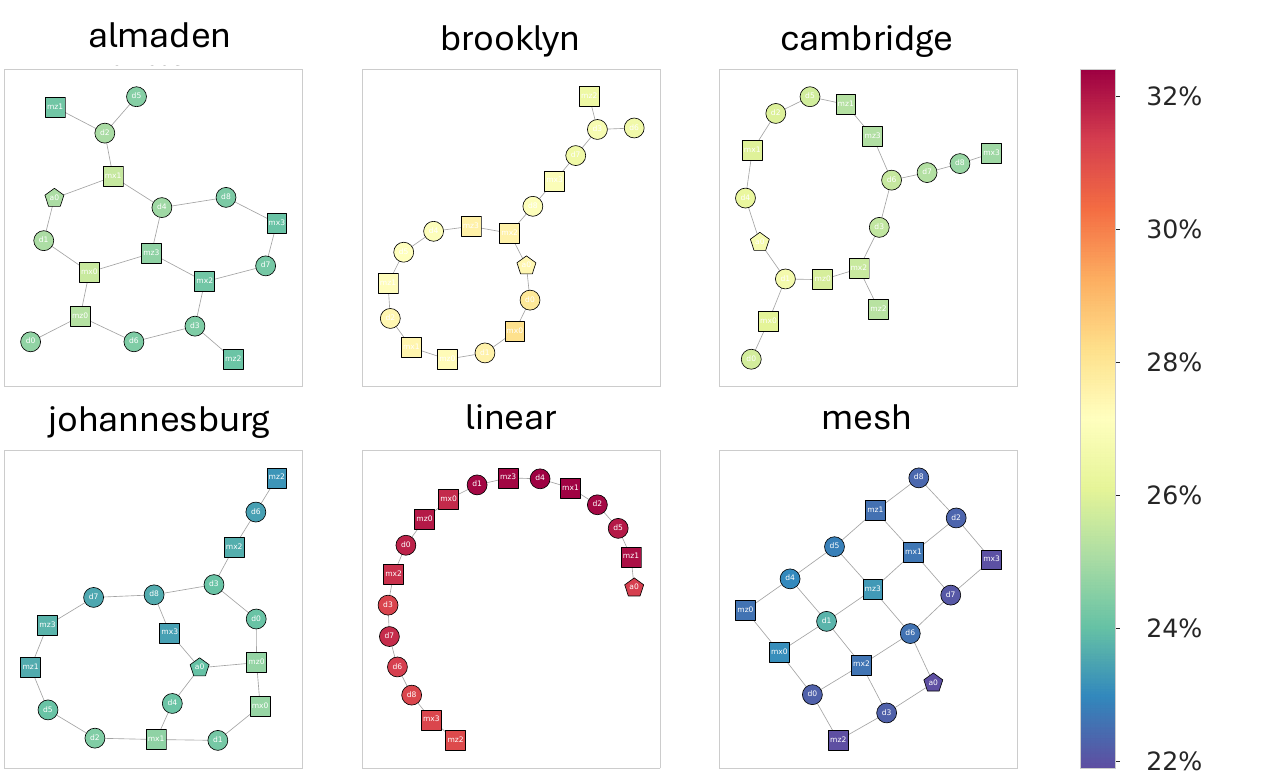}
    \caption{Distance-(3,3) XXZZ code.}
    \label{arch_xxzz}
\end{subfigure}
\caption{\textbf{Logical error rate by corrupted qubit on different architectures}. Plot of the logical error with respect to the root injection point. Each edge represents an integer distance of one between two nodes. Unused qubits in the original architectural graphs have been omitted. Note the color map scale difference between the left and the right plots.}
\label{architectural_plots}
\vspace{-10pt}
\end{figure*}

In Figure \ref{bar_chart_spread_depth}, we plot the post-decoding logical error for the \textit{distance-(15,1)} repetition code and the \textit{distance-(3,3)} XXZZ code only, as emblematic cases. 
The repetition code, when only one qubit is reset, shows a logical error rate of $\sim 17\%$, an absolute difference of $\sim 17\%$ with respect to the radiation-induced fault.
Increasing the number of qubits being reset monotonically increases the logical error rate for the correlated case, reaching a value of $\sim 25 \%$ when 15 qubits are erased.
As soon as more than half of the total number of qubits in the circuit are reset, which in the case of the distance-(15,1) repetition code is 15 qubits, the logical error rate reaches $\sim 80\%$, a value larger than the radiation-induced logical error rate of $\sim 34 \%$.

The XXZZ code, when a single qubit is reset, shows a logical error rate of $\sim 21\%$, which is almost one third of the logical error rate obtained with a radiation-induced fault.
As the number of erasure errors increases, the logical error rate worsens, and once ten qubits are corrupted, the logical error rate reaches $\sim 36 \%$.
Performance degrades again once at least 15 qubits are corrupted, reaching an error rate of $\sim 80\%$, a logical error rate that exceeds the one of a single-qubit radiation-induced fault.

A single radiation-induced fault, despite a rapid damping in intensity as distance grows (shown in Figure~\ref{spatial_damping_evolution}), has shown significantly more detrimental effects than multiple erasure faults.
This further highlights the danger of radiation-induced faults.

\obs{A single spatially correlated radiation-induced fault is more detrimental than multiple uncorrelated erasure events.}

Our analysis highlights that limiting the spatial spread of radiation-induced faults is crucial to guarantee the performance of surface codes. 
Hardware solutions that promise to prevent radiation events from spreading over the substrate \cite{Iaia2022, mcewen2024resisting, harrington2024synchronous, li2024direct}, can then improve the performance of surface codes.
Given that qubit isolation solutions will have a \textit{significant} impact on the production cost of quantum computers, it is fundamental to be able to test and validate their effect beforehand.
While completely removing the spatial spread of transient errors through complete isolation of each qubit on the substrate is an unreachable task, these techniques may prove to have a positive impact on the error correction capabilities of surface codes.
However, rarer events, that can independently corrupt multiple physical qubits at the same time, will still pose a threat to quantum reliability.

\obs{Limiting the spatial spread of the radiation-induced errors significantly improves the error correction capabilities of surface codes, increasing the threshold for the number of concurrent erasure errors that they can withstand.}

\subsection{Hardware architecture analysis}
\label{architectural_analysis}
Quantum circuits need to be transpiled following the architecture constraints of a quantum computer to be executed.
We ask ourselves: (1) Does the choice of the architectural connectivity graph impact the logical error rate? (2) Are specific qubits in the surface code more critical than others?
To answer these questions, we transpile the surface codes to various architectural connectivity graphs, injecting a single radiation-induced fault on each possible root injection point, and following its evolution over the temporal and spatial domain.
The transpilation process has been done with the default optimisation factor and without forcing the qubit positioning in the initial or final layouts.
The architecture graphs we tested include a $5\times6$ mesh, a linear graph and a few of the hardware connectivity patterns publicly available from the Qiskit library \cite{Qiskit}.
These latter graphs have been filtered according to the number of qubits required to represent the surface code and communication constraints.
We consider a one-to-one mapping between the architectural graph and each qubit's physical position on the quantum chip, and a constant weight of one on each edge.

We plot the results over each architectural graph.
The hue of each node represents the median logical error over the fault's duration obtained from a particle impact in that qubit.
The shape of the node represents the qubit's function in the surface code: data qubit nodes are enclosed by a circle, stabiliser qubit nodes are enclosed by a square and the ancilla qubit nodes are enclosed by a pentagon.
We show results for the \textit{distance-(11,1)} repetition code and the \textit{distance-(3,3)} XXZZ code. 

The repetition code, due to its 22 qubit size constraint, has been tested on these architecture graphs: linear, mesh ($5 \times 6$), Brooklyn, Cairo and Cambridge.
The linear architecture boasts the lowest median logical error range. It ranges from $\sim 15\%$ when injecting either the ancilla qubit or the lower indexed data and stabiliser qubits, to $\sim 17\%$ for the injections on higher indexed qubits.
The mesh architecture highlights a similar correlation, as the ancilla and the lower indexed data and stabilisation qubits show slightly larger median errors of $\sim 17\%$ than their higher indexed counterparts ($\sim 16\%$).
This correlation is true, albeit for slightly different error ranges, across all the other architecture graphs.
The Cairo architecture has the worst performance, reaching $23 \%$ error at its peak, and lower thresholds of $\sim 21.5 \%$.
The Brooklyn architecture shows a relatively stable median logical error rate of $\sim 19\%$, while the Cambridge architecture's median logical error ranges from $\sim 17 \%$ to $\sim 19\%$.
The linear and mesh architectures boast the lowest error rates since they better support the nearest neighbour interactions of the repetition code, which requires that stabiliser qubits are placed on nodes with degree $\ge 2$.

The XXZZ code, given the smaller size requirement of 18 qubits, has been tested on the following architecture graphs: complete, linear, mesh ($5 \times 4$), Almaden, Brooklyn, Cambridge and Johannesburg.
In this case, the mesh architecture sports the lowest median logical error, with a peak at $\sim 24.5\%$, and the lowest logical error at $\sim 22\%$.
The Johannesburg and Almaden architectures show higher error rates, ranging from $\sim 23\%$ to $\sim 26 \%$.
The Cambridge architecture has a higher logical error variation, which peaks on lower indexed qubits at $\sim 27\%$, whilst the Brooklyn architecture performs worse, at upwards of $\sim 28\%$ logical error rates and little variation with respect to the position of the corrupted qubit.
The linear architecture has a significantly worse performance when compared with the others.
This is due to the fact that the XXZZ code, being a rotated code, requires that the stabiliser qubits are placed on nodes with degree $\ge 4$, while the linear architecture has an average node degree $\lesssim 2$, thus introducing a large overhead in SWAP operations.

This analysis unveiled a correlation between the index of the qubit in the surface code, and the median logical error rate registered when that qubit acts as the locus of a transient fault event.
This matches the flow of information across qubits during the execution of the surface code's circuit.
Such behaviour can be explained by looking at the Directed Acyclic Graph (DAG) representation of the circuits analysed, which highlights the sequential dependence across multiple gate operations, as qubits get linked together by successive CNOT and SWAP gates.
When a single particle hits a qubit and spreads over the code, it will affect also the descendants in the DAG.
The lower error rate in higher indexed qubits is then just a matter of ordering convention in the temporal sequence of quantum gates applied.
\obs{Radiation-induced transient errors have a stronger impact on qubits that are used earlier in the sequence of gates in a quantum circuit.}
No circuit rewriting or reordering technique, such as the ones performed by a transpiler, can avoid this effect.
This is because the sequential dependence across gate operations in the analysed surface codes is intrinsic to their formalism.

Moreover, the choice of a quantum computer architecture significantly impacts the performance of surface codes according to their connectivity requirements, that is the average qubit degree required to represent the quantum circuit.
While the repetition code works exceptionally well when transpiled on a linear architecture, the XXZZ code suffers from a large overhead of SWAP operations, thus increasing the number of gate operations and the chance for a radiation-induced fault to propagate.

\obs{If the architecture graph of a quantum computer is sufficiently connected, there will be less communication overhead, preventing radiation-induced faults from spreading.}

\section{Conclusions and future works}
\label{conclusions_and_future_works}
In this paper, we have presented an analysis of the impact of radiation-induced faults on the efficacy of two classes of QEC codes, namely the repetition and XXZZ codes.
We answer three main research questions and found that
(RQ1) QEC surface codes cannot withstand the faults introduced by radiation. The extrapolated post-QEC logical error peaks we have observed reach $24\%$ and $54\%$.
(RQ2) We can indeed tune the QEC code to improve the reliability to radiation strikes. 
Our analysis shows that, given an equivalent number of physical qubits, the bit-flip correction codes are up to $10\%$ more effective against radiation than bit-phase correction codes.
Moreover, choosing properly the underlying hardware topology can further increase the radiation fault correction capability from 7\% to upwards of 10\%. These improvements do not introduce any additional overhead to the QEC.
(RQ3) We found that, to increase a surface code's resistance to radiation faults, bit-flip protection should be prioritised. Moreover, qubit charge wells are a promising solution to reduce the impact of transient faults, by preventing their spread. These insights can guide the design of future QEC codes able to cope also with radiation-induced errors.

Future research directions include 
the usage of the presented post-QEC logical error rates to perform post-QEC logical layer fault injection.
We intend to propagate the logical fault induced by radiation in the coded qubit status in quantum circuits.
The aim is to identify the critical logical shifts for a given circuit to further better tune the QEC correction capabilities.

\bibliographystyle{IEEEtran}
\bibliography{bibliography}

\begin{thebibliography}{10}
\providecommand{\url}[1]{#1}
\csname url@samestyle\endcsname
\providecommand{\newblock}{\relax}
\providecommand{\bibinfo}[2]{#2}
\providecommand{\BIBentrySTDinterwordspacing}{\spaceskip=0pt\relax}
\providecommand{\BIBentryALTinterwordstretchfactor}{4}
\providecommand{\BIBentryALTinterwordspacing}{\spaceskip=\fontdimen2\font plus
\BIBentryALTinterwordstretchfactor\fontdimen3\font minus \fontdimen4\font\relax}
\providecommand{\BIBforeignlanguage}[2]{{%
\expandafter\ifx\csname l@#1\endcsname\relax
\typeout{** WARNING: IEEEtran.bst: No hyphenation pattern has been}%
\typeout{** loaded for the language `#1'. Using the pattern for}%
\typeout{** the default language instead.}%
\else
\language=\csname l@#1\endcsname
\fi
#2}}
\providecommand{\BIBdecl}{\relax}
\BIBdecl

\bibitem{shor1994factorisation}
P.~Shor, ``Algorithms for quantum computation: discrete logarithms and factoring,'' in \emph{Proceedings 35th Annual Symposium on Foundations of Computer Science}, 1994, pp. 124--134.

\bibitem{1996Grover}
L.~K. {Grover}, ``{A fast quantum mechanical algorithm for database search},'' \emph{arXiv e-prints}, pp. quant--ph/9\,605\,043, May 1996.

\bibitem{preskill1998reliability}
\BIBentryALTinterwordspacing
J.~Preskill, ``Reliable quantum computers,'' \emph{Proceedings of the Royal Society of London. Series A: Mathematical, Physical and Engineering Sciences}, vol. 454, no. 1969, pp. 385--410, 1998. [Online]. Available: \url{https://royalsocietypublishing.org/doi/abs/10.1098/rspa.1998.0167}
\BIBentrySTDinterwordspacing

\bibitem{vanmeter2016}
R.~Van~Meter and S.~J. Devitt, ``The path to scalable distributed quantum computing,'' \emph{Computer}, vol.~49, no.~9, pp. 31--42, 2016.

\bibitem{sete2016}
E.~A. Sete, W.~J. Zeng, and C.~T. Rigetti, ``A functional architecture for scalable quantum computing,'' in \emph{2016 IEEE International Conference on Rebooting Computing (ICRC)}, 2016, pp. 1--6.

\bibitem{copsey2003}
D.~Copsey, M.~Oskin, F.~Impens, T.~Metodiev, A.~Cross, F.~Chong, I.~Chuang, and J.~Kubiatowicz, ``Toward a scalable, silicon-based quantum computing architecture,'' \emph{IEEE Journal of Selected Topics in Quantum Electronics}, vol.~9, no.~6, pp. 1552--1569, 2003.

\bibitem{Unruh1995}
\BIBentryALTinterwordspacing
W.~G. Unruh, ``Maintaining coherence in quantum computers,'' \emph{Phys. Rev. A}, vol.~51, pp. 992--997, Feb 1995. [Online]. Available: \url{https://link.aps.org/doi/10.1103/PhysRevA.51.992}
\BIBentrySTDinterwordspacing

\bibitem{divincenzo1999coherence}
\BIBentryALTinterwordspacing
D.~P. DiVincenzo and D.~Loss, ``Quantum computers and quantum coherence,'' \emph{Journal of Magnetism and Magnetic Materials}, vol. 200, no.~1, pp. 202--218, 1999. [Online]. Available: \url{https://www.sciencedirect.com/science/article/pii/S0304885399003157}
\BIBentrySTDinterwordspacing

\bibitem{Stassi2020}
\BIBentryALTinterwordspacing
R.~Stassi, M.~Cirio, and F.~Nori, ``Scalable quantum computer with superconducting circuits in the ultrastrong coupling regime,'' \emph{npj Quantum Information}, vol.~6, no.~1, p.~67, Aug 2020. [Online]. Available: \url{https://doi.org/10.1038/s41534-020-00294-x}
\BIBentrySTDinterwordspacing

\bibitem{Wang2022}
\BIBentryALTinterwordspacing
C.~Wang, X.~Li, H.~Xu, Z.~Li, J.~Wang, Z.~Yang, Z.~Mi, X.~Liang, T.~Su, C.~Yang, G.~Wang, W.~Wang, Y.~Li, M.~Chen, C.~Li, K.~Linghu, J.~Han, Y.~Zhang, Y.~Feng, Y.~Song, T.~Ma, J.~Zhang, R.~Wang, P.~Zhao, W.~Liu, G.~Xue, Y.~Jin, and H.~Yu, ``Towards practical quantum computers: transmon qubit with a lifetime approaching 0.5 milliseconds,'' \emph{npj Quantum Information}, vol.~8, no.~1, p.~3, Jan 2022. [Online]. Available: \url{https://doi.org/10.1038/s41534-021-00510-2}
\BIBentrySTDinterwordspacing

\bibitem{Somoroff2023}
\BIBentryALTinterwordspacing
A.~Somoroff, Q.~Ficheux, R.~A. Mencia, H.~Xiong, R.~Kuzmin, and V.~E. Manucharyan, ``Millisecond coherence in a superconducting qubit,'' \emph{Phys. Rev. Lett.}, vol. 130, p. 267001, Jun 2023. [Online]. Available: \url{https://link.aps.org/doi/10.1103/PhysRevLett.130.267001}
\BIBentrySTDinterwordspacing

\bibitem{Ghosh2013}
\BIBentryALTinterwordspacing
J.~Ghosh, A.~Galiautdinov, Z.~Zhou, A.~N. Korotkov, J.~M. Martinis, and M.~R. Geller, ``High-fidelity controlled-${\ensuremath{\sigma}}^{Z}$ gate for resonator-based superconducting quantum computers,'' \emph{Phys. Rev. A}, vol.~87, p. 022309, Feb 2013. [Online]. Available: \url{https://link.aps.org/doi/10.1103/PhysRevA.87.022309}
\BIBentrySTDinterwordspacing

\bibitem{Willsch2017}
\BIBentryALTinterwordspacing
D.~Willsch, M.~Nocon, F.~Jin, H.~De~Raedt, and K.~Michielsen, ``Gate-error analysis in simulations of quantum computers with transmon qubits,'' \emph{Phys. Rev. A}, vol.~96, p. 062302, Dec 2017. [Online]. Available: \url{https://link.aps.org/doi/10.1103/PhysRevA.96.062302}
\BIBentrySTDinterwordspacing

\bibitem{Rol2019}
\BIBentryALTinterwordspacing
M.~Rol, F.~Battistel, F.~Malinowski, C.~Bultink, B.~Tarasinski, R.~Vollmer, N.~Haider, N.~Muthusubramanian, A.~Bruno, B.~Terhal, and L.~DiCarlo, ``Fast, high-fidelity conditional-phase gate exploiting leakage interference in weakly anharmonic superconducting qubits,'' \emph{Physical Review Letters}, vol. 123, no.~12, Sep. 2019. [Online]. Available: \url{http://dx.doi.org/10.1103/PhysRevLett.123.120502}
\BIBentrySTDinterwordspacing

\bibitem{Kim2022}
\BIBentryALTinterwordspacing
Y.~Kim, A.~Morvan, L.~B. Nguyen, R.~K. Naik, C.~J{\"u}nger, L.~Chen, J.~M. Kreikebaum, D.~I. Santiago, and I.~Siddiqi, ``High-fidelity three-qubit itoffoli gate for fixed-frequency superconducting qubits,'' \emph{Nature Physics}, vol.~18, no.~7, pp. 783--788, Jul 2022. [Online]. Available: \url{https://doi.org/10.1038/s41567-022-01590-3}
\BIBentrySTDinterwordspacing

\bibitem{AbuGhanem2024}
\BIBentryALTinterwordspacing
M.~AbuGhanem and H.~Eleuch, ``Two-qubit entangling gates for superconducting quantum computers,'' \emph{Results in Physics}, vol.~56, p. 107236, 2024. [Online]. Available: \url{https://www.sciencedirect.com/science/article/pii/S221137972301029X}
\BIBentrySTDinterwordspacing

\bibitem{Gambetta2017}
\BIBentryALTinterwordspacing
J.~M. Gambetta, J.~M. Chow, and M.~Steffen, ``Building logical qubits in a superconducting quantum computing system,'' \emph{npj Quantum Information}, vol.~3, no.~1, p.~2, Jan 2017. [Online]. Available: \url{https://doi.org/10.1038/s41534-016-0004-0}
\BIBentrySTDinterwordspacing

\bibitem{Wendin2017}
\BIBentryALTinterwordspacing
G.~Wendin, ``Quantum information processing with superconducting circuits: a review,'' \emph{Reports on Progress in Physics}, vol.~80, no.~10, p. 106001, sep 2017. [Online]. Available: \url{https://dx.doi.org/10.1088/1361-6633/aa7e1a}
\BIBentrySTDinterwordspacing

\bibitem{Vischi2022}
\BIBentryALTinterwordspacing
M.~Vischi, L.~Ferialdi, A.~Trombettoni, and A.~Bassi, ``Possible limits on superconducting quantum computers from spontaneous wave-function collapse models,'' \emph{Phys. Rev. B}, vol. 106, p. 174506, Nov 2022. [Online]. Available: \url{https://link.aps.org/doi/10.1103/PhysRevB.106.174506}
\BIBentrySTDinterwordspacing

\bibitem{JavadiAbhari2017}
\BIBentryALTinterwordspacing
A.~Javadi-Abhari, P.~Gokhale, A.~Holmes, D.~Franklin, K.~R. Brown, M.~Martonosi, and F.~T. Chong, ``Optimized surface code communication in superconducting quantum computers,'' in \emph{Proceedings of the 50th Annual IEEE/ACM International Symposium on Microarchitecture}, ser. MICRO-50 '17.\hskip 1em plus 0.5em minus 0.4em\relax New York, NY, USA: Association for Computing Machinery, 2017, p. 692–705. [Online]. Available: \url{https://doi.org/10.1145/3123939.3123949}
\BIBentrySTDinterwordspacing

\bibitem{Andersen2020}
\BIBentryALTinterwordspacing
C.~K. Andersen, A.~Remm, S.~Lazar, S.~Krinner, N.~Lacroix, G.~J. Norris, M.~Gabureac, C.~Eichler, and A.~Wallraff, ``Repeated quantum error detection in a surface code,'' \emph{Nature Physics}, vol.~16, no.~8, pp. 875--880, Aug 2020. [Online]. Available: \url{https://doi.org/10.1038/s41567-020-0920-y}
\BIBentrySTDinterwordspacing

\bibitem{Krinner2022}
\BIBentryALTinterwordspacing
S.~Krinner, N.~Lacroix, A.~Remm, A.~Di~Paolo, E.~Genois, C.~Leroux, C.~Hellings, S.~Lazar, F.~Swiadek, J.~Herrmann, G.~J. Norris, C.~K. Andersen, M.~M{\"u}ller, A.~Blais, C.~Eichler, and A.~Wallraff, ``Realizing repeated quantum error correction in a distance-three surface code,'' \emph{Nature}, vol. 605, no. 7911, pp. 669--674, May 2022. [Online]. Available: \url{https://doi.org/10.1038/s41586-022-04566-8}
\BIBentrySTDinterwordspacing

\bibitem{Goto2023}
\BIBentryALTinterwordspacing
H.~Goto, Y.~Ho, and T.~Kanao, ``Measurement-free fault-tolerant logical-zero-state encoding of the distance-three nine-qubit surface code in a one-dimensional qubit array,'' \emph{Phys. Rev. Res.}, vol.~5, p. 043137, Nov 2023. [Online]. Available: \url{https://link.aps.org/doi/10.1103/PhysRevResearch.5.043137}
\BIBentrySTDinterwordspacing

\bibitem{Tiurev2023correcting}
\BIBentryALTinterwordspacing
K.~Tiurev, P.-J. H.~S. Derks, J.~Roffe, J.~Eisert, and J.-M. Reiner, ``Correcting non-independent and non-identically distributed errors with surface codes,'' \emph{{Quantum}}, vol.~7, p. 1123, Sep. 2023. [Online]. Available: \url{https://doi.org/10.22331/q-2023-09-26-1123}
\BIBentrySTDinterwordspacing

\bibitem{Katsuda2024}
\BIBentryALTinterwordspacing
M.~Katsuda, K.~Mitarai, and K.~Fujii, ``Simulation and performance analysis of quantum error correction with a rotated surface code under a realistic noise model,'' \emph{Phys. Rev. Res.}, vol.~6, p. 013024, Jan 2024. [Online]. Available: \url{https://link.aps.org/doi/10.1103/PhysRevResearch.6.013024}
\BIBentrySTDinterwordspacing

\bibitem{Steane1996}
\BIBentryALTinterwordspacing
A.~Steane, ``Multiple-particle interference and quantum error correction,'' \emph{Proceedings of the Royal Society of London. Series A: Mathematical, Physical and Engineering Sciences}, vol. 452, no. 1954, pp. 2551--2577, 1996. [Online]. Available: \url{https://royalsocietypublishing.org/doi/abs/10.1098/rspa.1996.0136}
\BIBentrySTDinterwordspacing

\bibitem{Bonilla_Ataides_2021}
\BIBentryALTinterwordspacing
J.~P. Bonilla~Ataides, D.~K. Tuckett, S.~D. Bartlett, S.~T. Flammia, and B.~J. Brown, ``The xzzx surface code,'' \emph{Nature Communications}, vol.~12, no.~1, Apr. 2021. [Online]. Available: \url{http://dx.doi.org/10.1038/s41467-021-22274-1}
\BIBentrySTDinterwordspacing

\bibitem{Barends2011}
R.~Barends, J.~Wenner, M.~Lenander, Y.~Chen, R.~C. Bialczak, J.~Kelly, E.~Lucero, P.~O’Malley, M.~Mariantoni, D.~Sank \emph{et~al.}, ``Minimizing quasiparticle generation from stray infrared light in superconducting quantum circuits,'' \emph{Applied Physics Letters}, vol.~99, no.~11, p. 113507, 2011.

\bibitem{radiation2011}
\BIBentryALTinterwordspacing
A.~D. Corcoles, J.~M. Chow, J.~M. Gambetta, C.~Rigetti, J.~R. Rozen, G.~A. Keefe, M.~Beth~Rothwell, M.~B. Ketchen, and M.~Steffen, ``Protecting superconducting qubits from radiation,'' \emph{Applied Physics Letters}, vol.~99, no.~18, p. 181906, 2011. [Online]. Available: \url{https://doi.org/10.1063/1.3658630}
\BIBentrySTDinterwordspacing

\bibitem{LossMechanisms2018}
\BIBentryALTinterwordspacing
L.~Gr\"unhaupt, N.~Maleeva, S.~T. Skacel, M.~Calvo, F.~Levy-Bertrand, A.~V. Ustinov, H.~Rotzinger, A.~Monfardini, G.~Catelani, and I.~M. Pop, ``Loss mechanisms and quasiparticle dynamics in superconducting microwave resonators made of thin-film granular aluminum,'' \emph{Phys. Rev. Lett.}, vol. 121, p. 117001, Sep 2018. [Online]. Available: \url{https://link.aps.org/doi/10.1103/PhysRevLett.121.117001}
\BIBentrySTDinterwordspacing

\bibitem{nature_rad}
\BIBentryALTinterwordspacing
A.~P. Veps{\"a}l{\"a}inen, A.~H. Karamlou, J.~L. Orrell, A.~S. Dogra, B.~Loer, F.~Vasconcelos, D.~K. Kim, A.~J. Melville, B.~M. Niedzielski, J.~L. Yoder, S.~Gustavsson, J.~A. Formaggio, B.~A. VanDevender, and W.~D. Oliver, ``Impact of ionizing radiation on superconducting qubit coherence,'' \emph{Nature}, vol. 584, no. 7822, pp. 551--556, 2020. [Online]. Available: \url{https://doi.org/10.1038/s41586-020-2619-8}
\BIBentrySTDinterwordspacing

\bibitem{Wilen2021}
\BIBentryALTinterwordspacing
C.~D. Wilen, S.~Abdullah, N.~Kurinsky, C.~Stanford, L.~Cardani, G.~d’Imperio, C.~Tomei, L.~Faoro, L.~Ioffe, C.~Liu \emph{et~al.}, ``Correlated charge noise and relaxation errors in superconducting qubits,'' \emph{Nature}, vol. 594, no. 7863, pp. 369--373, 2021. [Online]. Available: \url{https://doi.org/10.1038/s41586-021-03557-5}
\BIBentrySTDinterwordspacing

\bibitem{Cardani2021}
L.~Cardani, F.~Valenti, N.~Casali, G.~Catelani, T.~Charpentier, M.~Clemenza, I.~Colantoni, A.~Cruciani, G.~D’Imperio, L.~Gironi \emph{et~al.}, ``Reducing the impact of radioactivity on quantum circuits in a deep-underground facility,'' \emph{Nature communications}, vol.~12, no.~1, p. 2733, 2021.

\bibitem{Martinis2021}
\BIBentryALTinterwordspacing
J.~M. Martinis, ``Saving superconducting quantum processors from decay and correlated errors generated by gamma and cosmic rays,'' \emph{npj Quantum Information}, vol.~7, no.~1, p.~90, 2021. [Online]. Available: \url{https://doi.org/10.1038/s41534-021-00431-0}
\BIBentrySTDinterwordspacing

\bibitem{Chen2021}
\BIBentryALTinterwordspacing
Z.~Chen, K.~J. Satzinger, J.~Atalaya, A.~N. Korotkov, A.~Dunsworth, D.~Sank, C.~Quintana, M.~McEwen, R.~Barends, P.~V. Klimov, S.~Hong, C.~Jones, A.~Petukhov, D.~Kafri, S.~Demura, B.~Burkett, C.~Gidney, A.~G. Fowler, A.~Paler, H.~Putterman, I.~Aleiner, F.~Arute, K.~Arya, R.~Babbush, J.~C. Bardin, A.~Bengtsson, A.~Bourassa, M.~Broughton, B.~B. Buckley, D.~A. Buell, N.~Bushnell, B.~Chiaro, R.~Collins, W.~Courtney, A.~R. Derk, D.~Eppens, C.~Erickson, E.~Farhi, B.~Foxen, M.~Giustina, A.~Greene, J.~A. Gross, M.~P. Harrigan, S.~D. Harrington, J.~Hilton, A.~Ho, T.~Huang, W.~J. Huggins, L.~B. Ioffe, S.~V. Isakov, E.~Jeffrey, Z.~Jiang, K.~Kechedzhi, S.~Kim, A.~Kitaev, F.~Kostritsa, D.~Landhuis, P.~Laptev, E.~Lucero, O.~Martin, J.~R. McClean, T.~McCourt, X.~Mi, K.~C. Miao, M.~Mohseni, S.~Montazeri, W.~Mruczkiewicz, J.~Mutus, O.~Naaman, M.~Neeley, C.~Neill, M.~Newman, M.~Y. Niu, T.~E. O'Brien, A.~Opremcak, E.~Ostby, B.~Pat{\'o}, N.~Redd, P.~Roushan, N.~C. Rubin, V.~Shvarts, D.~Strain, M.~Szalay, M.~D. Trevithick,
  B.~Villalonga, T.~White, Z.~J. Yao, P.~Yeh, J.~Yoo, A.~Zalcman, H.~Neven, S.~Boixo, V.~Smelyanskiy, Y.~Chen, A.~Megrant, J.~Kelly, and G.~Q. AI, ``Exponential suppression of bit or phase errors with cyclic error correction,'' \emph{Nature}, vol. 595, no. 7867, pp. 383--387, Jul 2021. [Online]. Available: \url{https://doi.org/10.1038/s41586-021-03588-y}
\BIBentrySTDinterwordspacing

\bibitem{Acharya2023}
\BIBentryALTinterwordspacing
R.~Acharya, I.~Aleiner, R.~Allen, T.~I. Andersen, M.~Ansmann, F.~Arute, K.~Arya, A.~Asfaw, J.~Atalaya, R.~Babbush, D.~Bacon, J.~C. Bardin, J.~Basso, A.~Bengtsson, S.~Boixo, G.~Bortoli, A.~Bourassa, J.~Bovaird, L.~Brill, M.~Broughton, B.~B. Buckley, D.~A. Buell, T.~Burger, B.~Burkett, N.~Bushnell, Y.~Chen, Z.~Chen, B.~Chiaro, J.~Cogan, R.~Collins, P.~Conner, W.~Courtney, A.~L. Crook, B.~Curtin, D.~M. Debroy, A.~Del Toro~Barba, S.~Demura, A.~Dunsworth, D.~Eppens, C.~Erickson, L.~Faoro, E.~Farhi, R.~Fatemi, L.~Flores~Burgos, E.~Forati, A.~G. Fowler, B.~Foxen, W.~Giang, C.~Gidney, D.~Gilboa, M.~Giustina, A.~Grajales~Dau, J.~A. Gross, S.~Habegger, M.~C. Hamilton, M.~P. Harrigan, S.~D. Harrington, O.~Higgott, J.~Hilton, M.~Hoffmann, S.~Hong, T.~Huang, A.~Huff, W.~J. Huggins, L.~B. Ioffe, S.~V. Isakov, J.~Iveland, E.~Jeffrey, Z.~Jiang, C.~Jones, P.~Juhas, D.~Kafri, K.~Kechedzhi, J.~Kelly, T.~Khattar, M.~Khezri, M.~Kieferov{\'a}, S.~Kim, A.~Kitaev, P.~V. Klimov, A.~R. Klots, A.~N. Korotkov, F.~Kostritsa, J.~M.
  Kreikebaum, D.~Landhuis, P.~Laptev, K.-M. Lau, L.~Laws, J.~Lee, K.~Lee, B.~J. Lester, A.~Lill, W.~Liu, A.~Locharla, E.~Lucero, F.~D. Malone, J.~Marshall, O.~Martin, J.~R. McClean, T.~McCourt, M.~McEwen, A.~Megrant, B.~Meurer~Costa, X.~Mi, K.~C. Miao, M.~Mohseni, S.~Montazeri, A.~Morvan, E.~Mount, W.~Mruczkiewicz, O.~Naaman, M.~Neeley, C.~Neill, A.~Nersisyan, H.~Neven, M.~Newman, J.~H. Ng, A.~Nguyen, M.~Nguyen, M.~Y. Niu, T.~E. O'Brien, A.~Opremcak, J.~Platt, A.~Petukhov, R.~Potter, L.~P. Pryadko, C.~Quintana, P.~Roushan, N.~C. Rubin, N.~Saei, D.~Sank, K.~Sankaragomathi, K.~J. Satzinger, H.~F. Schurkus, C.~Schuster, M.~J. Shearn, A.~Shorter, V.~Shvarts, J.~Skruzny, V.~Smelyanskiy, W.~C. Smith, G.~Sterling, D.~Strain, M.~Szalay, A.~Torres, G.~Vidal, B.~Villalonga, C.~Vollgraff~Heidweiller, T.~White, C.~Xing, Z.~J. Yao, P.~Yeh, J.~Yoo, G.~Young, A.~Zalcman, Y.~Zhang, N.~Zhu, and G.~Q. AI, ``Suppressing quantum errors by scaling a surface code logical qubit,'' \emph{Nature}, vol. 614, no. 7949, pp. 676--681,
  Feb 2023. [Online]. Available: \url{https://doi.org/10.1038/s41586-022-05434-1}
\BIBentrySTDinterwordspacing

\bibitem{Oliveira2023neutrons}
D.~Oliveira, E.~Auden, and P.~Rech, ``Atmospheric neutron-induced fault generation and propagation in quantum bits and quantum circuits,'' \emph{IEEE Transactions on Nuclear Science}, vol.~70, no.~4, pp. 345--353, 2023.

\bibitem{Oliveira2017}
D.~A. G.~D. Oliveira, L.~L. Pilla, M.~Hanzich, V.~Fratin, F.~Fernandes, C.~Lunardi, J.~M. Cela, P.~O.~A. Navaux, L.~Carro, and P.~Rech, ``Radiation-induced error criticality in modern hpc parallel accelerators,'' in \emph{2017 IEEE International Symposium on High Performance Computer Architecture (HPCA)}, 2017, pp. 577--588.

\bibitem{asuqa_repo}
T.~B.D. (2024) Repository name. \url{To be disclosed after peer review}.

\bibitem{Iaia2022}
\BIBentryALTinterwordspacing
V.~Iaia, J.~Ku, A.~Ballard, C.~P. Larson, E.~Yelton, C.~H. Liu, S.~Patel, R.~McDermott, and B.~L.~T. Plourde, ``Phonon downconversion to suppress correlated errors in superconducting qubits,'' \emph{Nature Communications}, vol.~13, no.~1, Oct. 2022. [Online]. Available: \url{http://dx.doi.org/10.1038/s41467-022-33997-0}
\BIBentrySTDinterwordspacing

\bibitem{mcewen2024resisting}
M.~McEwen, K.~C. Miao, J.~Atalaya, A.~Bilmes, A.~Crook, J.~Bovaird, J.~M. Kreikebaum, N.~Zobrist, E.~Jeffrey, B.~Ying, A.~Bengtsson, H.-S. Chang, A.~Dunsworth, J.~Kelly, Y.~Zhang, E.~Forati, R.~Acharya, J.~Iveland, W.~Liu, S.~Kim, B.~Burkett, A.~Megrant, Y.~Chen, C.~Neill, D.~Sank, M.~Devoret, and A.~Opremcak, ``Resisting high-energy impact events through gap engineering in superconducting qubit arrays,'' 2024.

\bibitem{li2024direct}
X.-G. Li, J.-H. Wang, Y.-Y. Jiang, G.-M. Xue, X.-X. Cai, J.~Zhou, M.~Gong, Z.-F. Liu, S.-Y. Zheng, D.-K. Ma, M.~Chen, W.-J. Sun, S.~Yang, F.~Yan, Y.-R. Jin, X.-F. Ding, and H.-F. Yu, ``Direct evidence for cosmic-ray-induced correlated errors in superconducting qubit array,'' 2024.

\bibitem{Ravi2023}
\BIBentryALTinterwordspacing
G.~S. Ravi, J.~M. Baker, A.~Fayyazi, S.~F. Lin, A.~Javadi-Abhari, M.~Pedram, and F.~T. Chong, ``Better than worst-case decoding for quantum error correction,'' in \emph{Proceedings of the 28th ACM International Conference on Architectural Support for Programming Languages and Operating Systems, Volume 2}, ser. ASPLOS 2023.\hskip 1em plus 0.5em minus 0.4em\relax New York, NY, USA: Association for Computing Machinery, 2023, p. 88–102. [Online]. Available: \url{https://doi.org/10.1145/3575693.3575733}
\BIBentrySTDinterwordspacing

\bibitem{Sivak2023}
\BIBentryALTinterwordspacing
V.~V. Sivak, A.~Eickbusch, B.~Royer, S.~Singh, I.~Tsioutsios, S.~Ganjam, A.~Miano, B.~L. Brock, A.~Z. Ding, L.~Frunzio, S.~M. Girvin, R.~J. Schoelkopf, and M.~H. Devoret, ``Real-time quantum error correction beyond break-even,'' \emph{Nature}, vol. 616, no. 7955, pp. 50--55, Apr 2023. [Online]. Available: \url{https://doi.org/10.1038/s41586-023-05782-6}
\BIBentrySTDinterwordspacing

\bibitem{Preskill_2018}
\BIBentryALTinterwordspacing
J.~Preskill, ``Quantum computing in the nisq era and beyond,'' \emph{Quantum}, vol.~2, p.~79, Aug 2018. [Online]. Available: \url{http://dx.doi.org/10.22331/q-2018-08-06-79}
\BIBentrySTDinterwordspacing

\bibitem{Georgopoulos2021}
\BIBentryALTinterwordspacing
K.~Georgopoulos, C.~Emary, and P.~Zuliani, ``Modeling and simulating the noisy behavior of near-term quantum computers,'' \emph{Phys. Rev. A}, vol. 104, p. 062432, Dec 2021. [Online]. Available: \url{https://link.aps.org/doi/10.1103/PhysRevA.104.062432}
\BIBentrySTDinterwordspacing

\bibitem{harrington2024synchronous}
P.~M. Harrington, M.~Li, M.~Hays, W.~V.~D. Pontseele, D.~Mayer, H.~D. Pinckney, F.~Contipelli, M.~Gingras, B.~M. Niedzielski, H.~Stickler, J.~L. Yoder, M.~E. Schwartz, J.~A. Grover, K.~Serniak, W.~D. Oliver, and J.~A. Formaggio, ``Synchronous detection of cosmic rays and correlated errors in superconducting qubit arrays,'' 2024.

\bibitem{Baumann2005}
R.~Baumann, ``Soft errors in advanced computer systems,'' \emph{IEEE Design Test of Computers}, vol.~22, no.~3, pp. 258--266, May 2005.

\bibitem{mcewen2022resolving}
\BIBentryALTinterwordspacing
M.~McEwen, L.~Faoro, K.~Arya, A.~Dunsworth, T.~Huang, S.~Kim, B.~Burkett, A.~Fowler, F.~Arute, J.~C. Bardin, A.~Bengtsson, A.~Bilmes, B.~B. Buckley, N.~Bushnell, Z.~Chen, R.~Collins, S.~Demura, A.~R. Derk, C.~Erickson, M.~Giustina, S.~D. Harrington, S.~Hong, E.~Jeffrey, J.~Kelly, P.~V. Klimov, F.~Kostritsa, P.~Laptev, A.~Locharla, X.~Mi, K.~C. Miao, S.~Montazeri, J.~Mutus, O.~Naaman, M.~Neeley, C.~Neill, A.~Opremcak, C.~Quintana, N.~Redd, P.~Roushan, D.~Sank, K.~J. Satzinger, V.~Shvarts, T.~White, Z.~J. Yao, P.~Yeh, J.~Yoo, Y.~Chen, V.~Smelyanskiy, J.~M. Martinis, H.~Neven, A.~Megrant, L.~Ioffe, and R.~Barends, ``Resolving catastrophic error bursts from cosmic rays in large arrays of superconducting qubits,'' \emph{Nature Physics}, vol.~18, no.~1, pp. 107--111, Jan. 2022. [Online]. Available: \url{https://doi.org/10.1038/s41567-021-01432-8}
\BIBentrySTDinterwordspacing

\bibitem{hpca2015}
D.~Tiwari, S.~Gupta, J.~Rogers, D.~Maxwell, P.~Rech, S.~Vazhkudai, D.~Oliveira, D.~Londo, N.~DeBardeleben, P.~Navaux, L.~Carro, and A.~Bland, ``Understanding gpu errors on large-scale hpc systems and the implications for system design and operation,'' in \emph{2015 IEEE 21st International Symposium on High Performance Computer Architecture (HPCA)}, 2015, pp. 331--342.

\bibitem{selse2014}
P.~Rech, L.~Carro, N.~Wang, T.~Tsai, S.~K.~S. Hari, and S.~W. Keckler, ``{Measuring the Radiation Reliability of SRAM Structures in GPUS Designed for HPC},'' in \emph{IEEE 10th Workshop on Silicon Errors in Logic - System Effects (SELSE)}, 2014.

\bibitem{loer2024abatement}
B.~Loer, P.~M. Harrington, B.~Archambault, E.~Fuller, B.~Pierson, I.~Arnquist, K.~Harouaka, T.~D. Schlieder, D.~K. Kim, A.~J. Melville, B.~M. Niedzielski, J.~K. Yoder, K.~Serniak, W.~D. Oliver, J.~L. Orrell, R.~Bunker, B.~A. VanDevender, and M.~Warner, ``Abatement of ionizing radiation for superconducting quantum devices,'' 2024.

\bibitem{Wootters1982nocloning}
\BIBentryALTinterwordspacing
W.~K. Wootters and W.~H. Zurek, ``A single quantum cannot be cloned,'' \emph{Nature}, vol. 299, no. 5886, pp. 802--803, Oct 1982. [Online]. Available: \url{https://doi.org/10.1038/299802a0}
\BIBentrySTDinterwordspacing

\bibitem{Kitaev2003}
\BIBentryALTinterwordspacing
A.~Kitaev, ``Fault-tolerant quantum computation by anyons,'' \emph{Annals of Physics}, vol. 303, no.~1, p. 2–30, Jan. 2003. [Online]. Available: \url{http://dx.doi.org/10.1016/S0003-4916(02)00018-0}
\BIBentrySTDinterwordspacing

\bibitem{Brown2022}
B.~J. Brown, ``Conservation laws and quantum error correction: Toward a generalized matching decoder,'' \emph{IEEE BITS the Information Theory Magazine}, vol.~2, no.~3, pp. 5--19, 2022.

\bibitem{Marton2023}
\BIBentryALTinterwordspacing
A.~Márton and J.~K. Asbóth, ``Coherent errors and readout errors in the surface code,'' \emph{Quantum}, vol.~7, p. 1116, Sep. 2023. [Online]. Available: \url{http://dx.doi.org/10.22331/q-2023-09-21-1116}
\BIBentrySTDinterwordspacing

\bibitem{Vittal2023}
\BIBentryALTinterwordspacing
S.~Vittal, P.~Das, and M.~Qureshi, ``Astrea: Accurate quantum error-decoding via practical minimum-weight perfect-matching,'' in \emph{Proceedings of the 50th Annual International Symposium on Computer Architecture}, ser. ISCA '23.\hskip 1em plus 0.5em minus 0.4em\relax New York, NY, USA: Association for Computing Machinery, 2023. [Online]. Available: \url{https://doi.org/10.1145/3579371.3589037}
\BIBentrySTDinterwordspacing

\bibitem{Sundaresan2023}
\BIBentryALTinterwordspacing
N.~Sundaresan, T.~J. Yoder, Y.~Kim, M.~Li, E.~H. Chen, G.~Harper, T.~Thorbeck, A.~W. Cross, A.~D. C{\'o}rcoles, and M.~Takita, ``Demonstrating multi-round subsystem quantum error correction using matching and maximum likelihood decoders,'' \emph{Nature Communications}, vol.~14, no.~1, p. 2852, May 2023. [Online]. Available: \url{https://doi.org/10.1038/s41467-023-38247-5}
\BIBentrySTDinterwordspacing

\bibitem{higgott2023sparse}
O.~Higgott and C.~Gidney, ``Sparse blossom: correcting a million errors per core second with minimum-weight matching,'' \emph{arXiv preprint arXiv:2303.15933}, 2023.

\bibitem{Bravyi2014}
\BIBentryALTinterwordspacing
S.~Bravyi, M.~Suchara, and A.~Vargo, ``Efficient algorithms for maximum likelihood decoding in the surface code,'' \emph{Phys. Rev. A}, vol.~90, p. 032326, Sep 2014. [Online]. Available: \url{https://link.aps.org/doi/10.1103/PhysRevA.90.032326}
\BIBentrySTDinterwordspacing

\bibitem{Old2023}
\BIBentryALTinterwordspacing
J.~Old and M.~Rispler, ``Generalized belief propagation algorithms for decoding of surface codes,'' \emph{Quantum}, vol.~7, p. 1037, Jun. 2023. [Online]. Available: \url{http://dx.doi.org/10.22331/q-2023-06-07-1037}
\BIBentrySTDinterwordspacing

\bibitem{Delfosse2021}
\BIBentryALTinterwordspacing
N.~Delfosse and N.~H. Nickerson, ``Almost-linear time decoding algorithm for topological codes,'' \emph{Quantum}, vol.~5, p. 595, Dec. 2021. [Online]. Available: \url{http://dx.doi.org/10.22331/q-2021-12-02-595}
\BIBentrySTDinterwordspacing

\bibitem{Varsamopoulos2018}
\BIBentryALTinterwordspacing
S.~Varsamopoulos, B.~Criger, and K.~Bertels, ``Decoding small surface codes with feedforward neural networks,'' \emph{Quantum Science and Technology}, vol.~3, no.~1, p. 015004, nov 2017. [Online]. Available: \url{https://dx.doi.org/10.1088/2058-9565/aa955a}
\BIBentrySTDinterwordspacing

\bibitem{ganjam2023surpassing}
S.~Ganjam, Y.~Wang, Y.~Lu, A.~Banerjee, C.~U. Lei, L.~Krayzman, K.~Kisslinger, C.~Zhou, R.~Li, Y.~Jia, M.~Liu, L.~Frunzio, and R.~J. Schoelkopf, ``Surpassing millisecond coherence times in on-chip superconducting quantum memories by optimizing materials, processes, and circuit design,'' 2023.

\bibitem{yelton2024modeling}
E.~Yelton, C.~P. Larson, V.~Iaia, K.~Dodge, G.~L. Magna, P.~G. Baity, I.~V. Pechenezhskiy, R.~McDermott, N.~Kurinsky, G.~Catelani, and B.~L.~T. Plourde, ``Modeling phonon-mediated quasiparticle poisoning in superconducting qubit arrays,'' 2024.

\bibitem{lin2024codesign}
S.~F. Lin, J.~Viszlai, K.~N. Smith, G.~S. Ravi, C.~Yuan, F.~T. Chong, and B.~J. Brown, ``Codesign of quantum error-correcting codes and modular chiplets in the presence of defects,'' 2024.

\bibitem{xiao2024exact}
Y.~Xiao, B.~Srivastava, and M.~Granath, ``Exact results on finite size corrections for surface codes tailored to biased noise,'' 2024.

\bibitem{qtcodes}
S.~Jha, A.~Ebrahimi, and J.~Gong. (2022) Qiskit topological codes. \url{https://github.com/yaleqc/qtcodes}.

\bibitem{Wootton2020}
\BIBentryALTinterwordspacing
J.~R. Wootton, ``Benchmarking near-term devices with quantum error correction,'' \emph{Quantum Science and Technology}, vol.~5, no.~4, p. 044004, Jul. 2020. [Online]. Available: \url{http://dx.doi.org/10.1088/2058-9565/aba038}
\BIBentrySTDinterwordspacing

\bibitem{BonillaAtaides2021}
\BIBentryALTinterwordspacing
J.~P. Bonilla~Ataides, D.~K. Tuckett, S.~D. Bartlett, S.~T. Flammia, and B.~J. Brown, ``The xzzx surface code,'' \emph{Nature Communications}, vol.~12, no.~1, p. 2172, Apr 2021. [Online]. Available: \url{https://doi.org/10.1038/s41467-021-22274-1}
\BIBentrySTDinterwordspacing

\bibitem{forlivesi2023logical}
D.~Forlivesi, L.~Valentini, and M.~Chiani, ``Logical error rates of xzzx and rotated quantum surface codes,'' 2023.

\bibitem{Qiskit}
e.~a. MD~SAJID~ANIS, ``Qiskit: An open-source framework for quantum computing,'' 2021.

\end{thebibliography}

\end{document}